\documentclass[11pt]{article}
\usepackage{amssymb,amsmath,graphicx}
\usepackage{amsfonts}
\usepackage[caption=false]{subfig}
\usepackage{epsfig,latexsym,graphicx}
\usepackage[sort,numbers,compress]{natbib}
\newtheorem{definition}{Definition}[section]
\usepackage{setspace}
\begin{document}

\title{Robust Wald-Type Tests under Random Censoring \\with Applications to Clinical Trial Analyses}

\author{
	Abhik Ghosh$^{1}$, Ayanendranath Basu$^{1 \ast}$ and Leandro Pardo$^2$
\\ $^1$ Indian Statistical Institute, Kolkata, India
\\ $^2$ Complutense University, Madrid, Spain
\\ $^{*}$Corresponding author. Email: {\it ayanbasu@isical.ac.in}
}
\date{}
\maketitle


\begin{abstract}
Randomly censored survival data are frequently encountered in applied sciences including biomedical or reliability applications and clinical trial analyses. 
Testing the significance of statistical hypotheses is crucial in such analyses to get conclusive inference
but the existing likelihood based tests, under a fully parametric model, are extremely non-robust against outliers in the data.
Although, there exists a few robust parameter estimators (e.g., M-estimators and minimum density power divergence  estimators)
given randomly censored data,  there is hardly any robust testing procedure available in the literature in this context. 
One of the major difficulties in this context is the construction of a suitable consistent estimator of 
the asymptotic variance of M-estimators; the latter is a function of the unknown censoring distribution. 
In this paper, we take the first step in this direction by proposing a consistent estimator of 
asymptotic variance of the  M-estimators based on randomly censored data without any assumption on the form of the censoring scheme. 
We then describe and study a class of robust Wald-type tests for parametric statistical hypothesis, 
both simple as well as composite, under such set-up, along with their general asymptotic and robustness properties.
Robust tests for comparing two independent randomly censored samples and robust tests against one sided alternatives are also discussed.
Their advantages and usefulness are demonstrated for the tests based on the minimum density power divergence estimators
with specific attention to clinical trial analyses.
\end{abstract}



\section{Introduction}\label{SEC:intro}

Randomly censored survival data are frequently encountered in different applied sciences including biomedical  and reliability applications. 
Some observed life-times are often seen to be right censored 
since the subject may still be alive at the end of study period or may have been lost to follow-up within the study period.
In most clinical trials, the censoring mechanism of patients depends on several unknown quantities 
and hence is generally assumed to be random and independent of the main response (e.g., patients lifetime).
Mathematically, $n$ subjects have life-time ($X$) measures denoted by $X_1, \ldots, X_n$, 
which are independent and identically distributed (i.i.d.) with distribution $G_{X}$.
However, due to right censoring, we only observe 
$Z_i = \min \left(X_i, C_i\right)$ and $\delta_i = I(X_i \leq C_i)$, for $i=1, \ldots, n$,
where $I(A)$ denotes the indicator function of the event $A$
and $C_1, \ldots, C_n$ denote $n$ i.i.d.~realizations of the censoring variable $C$ having distribution $G_C$.
We assume that $C$ is independent of $X$ and wish to do inference about the distribution of $X$ 
based on the observed data $\{Z_i, \delta_i\}_{i=1, \ldots, n}$.

\begin{figure}[h!t]
	\centering
\subfloat[Veteran Lung Cancer Trial]{
	\includegraphics[width=0.4\textwidth]{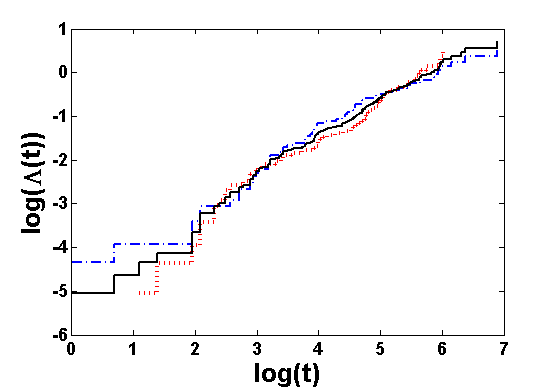}
	\label{FIG:loglogCumHaz_Veteran}}
	~ 
	\subfloat[Small-Cell Lung Cancer Trial]{
		\includegraphics[width=0.4\textwidth]{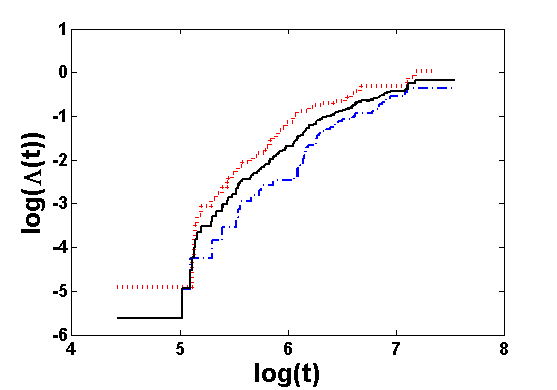}
		\label{FIG:loglogCumHaz_SmallCell}}
\\
	\subfloat[Breast and Ovarian Cancer Trial]{
	\includegraphics[width=0.4\textwidth]{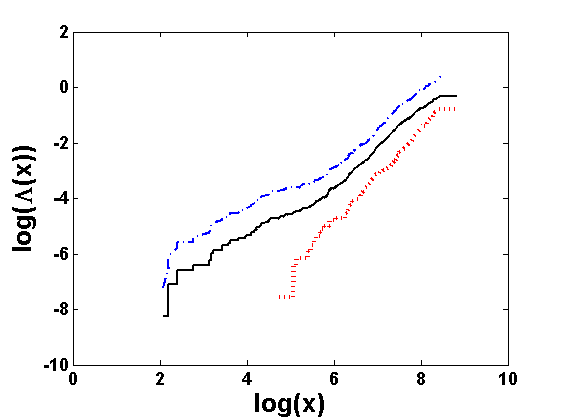}
	\label{FIG:loglogCumHaz_SmallCellwo}}
~ 
	\subfloat[Gastric Carcinoma Trial]{
	\includegraphics[width=0.4\textwidth]{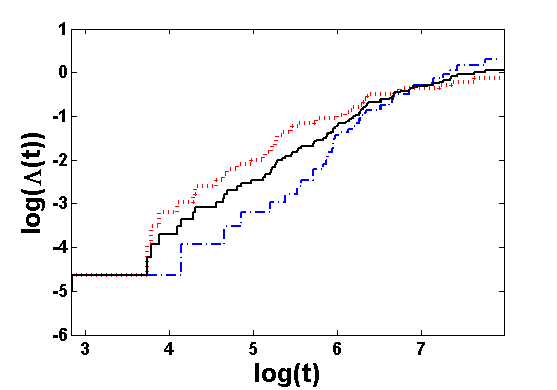}
	\label{FIG:loglogCumHaz_Gastric}}
	\caption{Empirical Cumulative hazards (in log-log scale) for different two-arm clinical trial data 
	(red dashed: Arm A, blue dash-dotted: Arm B, solid black: Combined data)}
	\label{FIG:CumHarad}
\end{figure}

Several non-parametric and semi-parametric inference procedures are available in the literature to handle such data
based on the Kaplan-Meier product-limit (KMPL) estimator of $G_X$ \citep{Kaplan/Meier:1958}, given by 
\begin{eqnarray}
\widehat{G_X}(x) = 1 - \prod_{i=1}^{n}\left[1 - \frac{\delta_{[i, n]}}{n-i+1}\right]^{I(Z_{(i,n)}\leq x)},
\label{EQ:KMPL}
\end{eqnarray}
where $Z_{(i, n)}$ denote the $i$-th order statistic in $\{Z_1, \cdots, Z_n\}$ and 
$\delta_{[i, n]}$ is the value of the corresponding $\delta$ ($i$-th concomitant).
Under the presence of random censoring, this popular estimator is the non-parametric maximum likelihood estimator (MLE) 
of the distribution function $G_X$ and enjoys many optimality properties. 
However, such non-parametric (or related semi-parametric) inference about $X$ is generally much less efficient compared to 
the procedures  implemented under some fully parametric distributional assumptions. 
See Chapter 8 of \cite{Hosmer/etc:2008} for advantages of such fully parametric modeling;
these include greater efficiency, easily interpretable parameter estimates and possibility of predictions from fitted models.
In medical sciences, such parametric assumptions can often be made from prior knowledge about the field 
or from similar experiments done in past (with same disease/drugs). 
From an empirical exploration of the data, we can also get an idea about the underlying parametric model;
for example, a linear plot of the empirical estimate of cumulative hazard, [$ -\log(1 - \widehat{G_X}(x))$],
over lifetime $x$ in a log-log scale indicates the suitability of the  Weibull family to be a possible candidate for modeling the underlying data
(see Figure \ref{FIG:CumHarad} for some clinical trial examples).

Classical fully parametric inference procedures are mainly based on the maximum likelihood approach;
see \cite{Cox/Oakes:1984,Crowder/etc:1991,Collett:2003,Lawless:2003,Klein/Moeschberger:2003} among many others.  
However, as in several other types of data, the maximum likelihood methods for censored data 
are also highly non-robust with respect to outliers.
Since outliers are not uncommon in real-life applications, suitable robust procedures having good model efficiency are always very useful. 
However, the robustness issue under survival data has been ignored in the literature for a long time and even now has only got some scattered attention.
A family of general M-estimators under randomly censored data has been introduced by \cite{Wang:1999}.
A particular M-estimator based on the density power divergence (DPD) of \cite{Basu/etc:1998} under a fully parametric set-up 
has been investigated by \cite{Basu/etc:2006}, exhibiting  robust performances with high efficiency. 
These procedures have recently been extended to a general regression-type set-up with 
stochastic covariates and randomly censored response \citep{Ghosh/Basu:2017}.
Robust estimators under semi-parametric accelerated failure time (AFT) models have been
developed by \cite{Zhou:2010,Locatelli/etc:2011,Wang/etc:2015}.
However, the current literature seems to offer very little in terms of robust testing of statistical hypotheses under randomly censored data, 
although the issue  is of high practical importance in real-life problems. 
To further motivate the sheer need of a robust testing approach, in particular for medical statistics, 
let us consider a few clinical trial data examples.

\begin{table}[h]
	\centering
	\caption{Parameter estimates for the Weibull model fitted to the four clinical trial datasets}
	\begin{tabular}{lr| ccc} \hline
		Clinical Trial Dataset		& Arm & 		\multicolumn{3}{c}{Estimate of $(a, b)$}\\
		&&MLE &	MDPDE at $\alpha=0.5$	&	MDPDE at $\alpha=1$	\\\hline
		Veteran Lung Cancer 	& Arm A & (123, 0.99) & (125, 0.96) & (122, 0.97) \\		
		& Arm B & (118, 0.76) & (95, 0.92) & (85, 0.99) \\		
		& Overall & (121, 0.85) & (111, 0.93) & (103, 0.97) \\		
		\hline
		Small-Cell Lung Cancer 	
		& Arm A & (987, 1.53) & (918, 1.51) & (731, 1.83) \\		
		& Arm B & (665, 1.36) & (480, 1.81) & (428, 2.10) \\		
		& Overall & (830, 1.38) & (724, 1.47) & (584, 1.78) \\		
		\hline
		Breast \& Ovarian Cancer	
		& Arm A & (4621, 1.79) & (4807, 1.62) & (4881, 1.54) \\		
		& Arm B & (1769, 1.28) & (1766, 1.34) & (1735, 1.48) \\		
		& Overall & (3075, 1.26) & (3030, 1.24) & (2798, 1.32) \\		
		\hline
		Gastric Carcinoma 	& Arm A & (760, 1.15) & (660, 1.48) & (600, 1.75) \\		
		& Arm B & (628, 0.79) & (376, 1.11) & (359, 1.18) \\		
		& Overall & (700, 0.92) & (593, 1.07) & (543, 1.19) \\		
		\hline
	\end{tabular}
	\label{TAB:MDPDE_4data}
\end{table}

\bigskip\noindent
\textbf{Motivating Examples:}\\
We focus on four clinical trials each having two groups denoted by Arm A and Arm B, respectively,
related to lung cancer, Breast and Ovarian Cancer (BOC) and Gastric Carcinoma; 
the two arms denote two groups of patients receiving two different treatments (one of which may be the control).
The details of these trials are provided in Section \ref{SEC:numerical};
but the empirical estimates of cumulative hazards based on the the right-censored lifetime data from these four trials 
are plotted in the log-log scale in Figure \ref{FIG:CumHarad} for both the arms separately as well as for the combined data. 
Note that, except at the extreme ends, either due to unrealistic early failures or outlying observations, 
the major part of these plots can be fitted well by a straight line which led us to the Weibull model having 
cumulative hazard $\Lambda(x) = a^b x^b$ with $a, b>0$ being the scale and the shape parameters, respectively.
So, we can analyze these clinical trials more efficiently by fitting a Weibull model to the data;
the inference will be based on the estimated values of the parameter $\boldsymbol{\theta} = (a, b)^T$ for each arm separately 
followed by suitable comparisons. 
Table \ref{TAB:MDPDE_4data} presents the MLE of these parameters for all four trials
along with the two robust estimators from the family of MDPDEs of \cite{Basu/etc:2006} 
under the Weibull model (See Section \ref{SEC:MDPDE} for definitions).
These robust estimators, depending on the value of tuning parameter $\alpha$, 
are slightly less efficient under pure data but extremely robust against data contamination.  
Note that, for all the trials, the MLE and the robust estimators differ from each other to a certain extent, sometimes substantially. 
For the inference to be conclusive, the practitioner would want to know
 whether these differences between the estimators are indeed statistically significant due to presence of outlying observations
or just due to sampling fluctuations. 
This depends on many factors like sample size, variance of the estimators, etc.,
and has to be rigorously confirmed through tests of hypothesis. There could be several important hypotheses to be tested in such situations.
Is there a small proportion of outliers which, while of interest on their own, influence the final decision by their presence? 
Do the model parameters for the two arms differ significantly only due to the presence of outliers 
(or are forced to look similar by the outliers masking a true difference)? 
Can a real improvement of treatment over control be hidden by a small set outliers? 
When we are suspicious about the presence of few extreme observations in data, which is quite common in practice, 
the inference obtained by using the MLE based tests to answer these questions can be often misleading. 
For example, in the Veteran Lung Cancer trial, the MLE of the shape parameter ($b$) differs significantly between arms
and the associated Wald test rejects the hypothesis of their equality in two arms ($b_{A}=b_{B}$) at standard significance levels.
However, their robust estimates are substantially closer, raising the suspicion that 
the earlier observed difference might be due to a small proportion of discordant observations. 
But, to prove this intuitive conclusion rigorously, 
we definitely need an appropriate testing procedures to robustly compare the two arms.
Similar erroneous inference can also be observed for testing the hypothesis that the shape parameter ($b$) for arm B data of 
the BOC trial is 1.45 using the likelihood based methods; 
such a test rejects this hypothesis based on the full data but accepts it after removing the outliers from the data.
Instead of the two-step outlier identification methods, a robust test of hypothesis procedure based 
on appropriate robust estimators (like MDPDE) will always be extremely effective and efficient in such cases; 
only having robust estimators itself will not suffice for a complete practical conclusion. 
The MDPDEs of $b$ for arm B data of the BOC trial, as reported in Table \ref{TAB:MDPDE_4data}, 
are very close to the null value 1.45, but we need to check if they are significantly close,
and that can only be done through a proper statistical test.
\hfill{$\square$}

Therefore,  nonavailability of robust test procedures under random censoring predominantly limits  the use of robust estimators too
and leads to the false conclusion in presence of outliers while using the only available likelihood based tests.
The primary obstacle in constructing robust tests has been the unknown censoring distribution
which is always involved in the asymptotic distribution of the robust estimators.
Hence, consistently estimating the asymptotic variance of the robust estimators based on randomly censored data has been the main challenge.
In this paper, we address this important problem by developing a class of robust Wald-type tests under randomly censored data.
For this purpose we develop a consistent estimator of the asymptotic variance of the M-estimators with an unknown censoring scheme. 
As a particular case, we study the advantages of the MDPDE based Wald-type tests 
under the fully parametric set-up; high efficiency of the MDPDE translates to high power for the corresponding tests.
%
%
We describe the procedure for testing parametric hypothesis under both one and two sample problems.
Besides demonstrating their robustness properties, 
we illustrate the usefulness of the proposed Wald-type tests in robust inference and comparison of clinical trial groups. 
All real datasets introduced as motivating examples are further studied in details using our proposal
along with two additional interesting datasets. 

\section{Robust Estimators under Random Censoring}\label{SEC:Est}

\subsection{General M-estimators}\label{SEC:Mest}

Although a few early approaches were previously available \citep{Reid:1981,Hjort:1985,Oakes:1986,James:1986,Lai/Ying:1994}, 
a formal M-estimation theory for randomly censored data was first developed by \cite{Wang:1995,Wang:1999}
with asymptotic results under simpler verifiable conditions.
Under the notation of Section \ref{SEC:intro}, consider the problem of estimating a parameter 
$\boldsymbol{\boldsymbol\theta}=\boldsymbol{\boldsymbol\theta}(G_X)$ from a parameter space 
$\Theta\subseteq \mathbb{R}^p$ ($p\geq1$).
In the fully parametric set-up, we model $G_X$ by a parametric family of distributions,
say $\mathcal{F}=\{F_{\boldsymbol{\boldsymbol\theta}} : \boldsymbol{\boldsymbol\theta}\in \Theta\subseteq\mathbb{R}^p\}$, 
with $\boldsymbol{\boldsymbol\theta}$ being the parameter of interest. 
Let $\boldsymbol{\theta}_0$ denote the true value of the parameter.

\begin{definition}	\label{DEF:M-est}
	Given a function $\boldsymbol\psi(x; \boldsymbol{\boldsymbol\theta}):~\mathbb{R}\times \Theta \mapsto \mathbb{R}^p$,
	the corresponding M-estimator of $\boldsymbol{\boldsymbol\theta}$ under random censoring is defined as 
	a solution to the estimating equation 
	$\int  \boldsymbol\psi(x; \boldsymbol{\boldsymbol\theta}) d\widehat{G_X}(x) = 0$,
	where $\widehat{G_X}(\cdot)$ is the KMPL estimator of $G_X$ given by (\ref{EQ:KMPL}), 
	and the $\boldsymbol\psi$-function satisfies
	\begin{equation}
	\int  \boldsymbol\psi(x; {\boldsymbol\theta}_0) d{G_X}(x) = 0.
	\label{EQ:est-eqn-psiX} 
	\end{equation}
\end{definition}
Under Assumption (\ref{EQ:est-eqn-psiX}), M-estimators are Fisher consistent 
and the associated estimating equation  is unbiased. 
In practice, while solving it numerically,
we may face the problem of multiple roots and some additional techniques are required in such cases.

Whenever $\boldsymbol\psi(x;{\boldsymbol\theta})$ is continuous and bounded  in ${\boldsymbol\theta}$,
there exists a strongly consistent (for ${\boldsymbol\theta}_0$) 
sequence of M-estimators and any sequence of M-estimators converges to ${\boldsymbol\theta}_0$ with probability one
\citep[][Theorem 3]{Wang:1999} under the assumptions listed in Appendix \ref{APP:Assumptions}. 
Further, if we define the random variables $Z=\min(X,C)$ and $\delta=I(X\leq C)$ and 
the sub-distribution functions $G_{Z,0}(z) = P(Z\leq z, \delta=0)$ and  $G_{Z,1}(z) = P(Z\leq z, \delta=1)$,
then the distribution of $Z$ is $G_Z=G_{Z,0}+G_{Z,1}$. 
For any real valued function $\phi(x)$, let us denote
$
U_{Z,\delta}(\phi)=\phi(Z)\gamma_0(Z)\delta + \gamma_{1}(Z;\phi)(1-\delta) - \gamma_{2}(Z;\phi) - \int \phi dG_X,
$
where
\begin{eqnarray}
\gamma_0(x) &=& \exp\left\{\int \frac{I(z<x)dG_{Z,0}(z)}{1-G_{Z}(z)}\right\},
~~
\gamma_1(x;\phi) =  \int \frac{I(z>x)\phi(z)\gamma_0(z)}{1-G_{Z}(x)}dG_{Z,1}(z),\nonumber
\end{eqnarray}
\begin{eqnarray}
\mbox{and }~\gamma_2(x;\phi) &=& 	\int \phi(z)\gamma_0(z)\gamma(\min\{x,z\})dG_{Z,1}(z),
~\mbox{ with }~\gamma(x) = \int \frac{I(z<x)dG_{Z,0}(z)}{\left[1-G_{Z}(z)\right]^2},\nonumber
\end{eqnarray}
and define the $p\times p$ matrices 
$\boldsymbol\Lambda(\boldsymbol\psi;{\boldsymbol\theta}) = \int \frac{\partial}{\partial{\boldsymbol\theta}} \boldsymbol\psi(x;{\boldsymbol\theta})dG_X(x)$
and
\begin{eqnarray}
\boldsymbol{C}(\boldsymbol\psi;{\boldsymbol\theta}) &=& E\left[(U_{Z,\delta}(\psi_1(\cdot;{\boldsymbol\theta})), \cdots, U_{Z,\delta}(\psi_p(\cdot;{\boldsymbol\theta})))^T
(U_{Z,\delta}(\psi_1(\cdot;{\boldsymbol\theta})), \cdots, U_{Z,\delta}(\psi_p(\cdot;{\boldsymbol\theta})))\right].\nonumber
\end{eqnarray}
Then, assuming that $\boldsymbol\psi(x;{\boldsymbol\theta})$ 
is differentiable in ${\boldsymbol\theta}$ in a neighborhood of true ${\boldsymbol\theta}_0$, 
along with Assumptions (A1)--(A6) listed in Appendix \ref{APP:Assumptions},
it has been shown that \cite[][Theorem 5]{Wang:1999} 
$\sqrt{n} \left(\widehat{\boldsymbol\theta}_n - {\boldsymbol\theta}_0 \right)$ 
is asymptotically normal with mean $\boldsymbol{0}_p$, the $p$-vector of zeros, and covariance matrix
$\boldsymbol\Sigma(\boldsymbol\psi; {\boldsymbol\theta}_0) = \boldsymbol\Lambda(\boldsymbol\psi;{\boldsymbol\theta}_0)^{-1}
\boldsymbol{C}(\boldsymbol\psi;{\boldsymbol\theta}_0)\boldsymbol\Lambda(\boldsymbol\psi;{\boldsymbol\theta}_0)^{-1}$,
for any consistent sequence $\{\widehat{\boldsymbol\theta}_n\}$ of corresponding M-estimators.

\subsection{The Minimum Density Power Divergence Estimator}\label{SEC:MDPDE}

A particular fully parametric M-estimator with high efficiency has been proposed by \cite{Basu/etc:2006} based on the DPD measure.
The DPD measure between two densities $g$ and $f$, with respect to a common dominating measure, is defined as
\begin{equation}
\label{EQ:dpd}
d_\alpha(g,f) = \displaystyle \left\{\begin{array}{ll}
\displaystyle \int  \left[f^{1+\alpha} - \left(1 + \frac{1}{\alpha}\right)  f^\alpha g + 
\frac{1}{\alpha} g^{1+\alpha}\right], & {\rm for} ~\alpha > 0,\\
\displaystyle \int g \log(g/f), & {\rm for} ~\alpha = 0.  
\end{array}\right.
\end{equation}
When we have $n$ i.i.d.~observations $Y_1, \ldots, Y_n$, having true density $g$, 
modeled by the parametric densities $\{f_{\boldsymbol{\boldsymbol\theta}} : {\boldsymbol\theta} \in \Theta \subset \mathbb{R}^p\}$,
the MDPDE of ${\boldsymbol\theta}$ is defined as the minimizer of the DPD between the data and $f_{\boldsymbol{\boldsymbol\theta}}$
with respect to ${\boldsymbol\theta}$, or equivalently as the minimizer of 
\begin{eqnarray}
\int f_{\boldsymbol{\boldsymbol\theta}}^{1+\alpha}(y)dy - \frac{1+\alpha}{\alpha} \int  f_{\boldsymbol{\boldsymbol\theta}}^{\alpha}(y)dG_n(y) 
=\int f_{\boldsymbol{\boldsymbol\theta}}^{1+\alpha}(y)dy - \frac{1+\alpha}{\alpha} \frac{1}{n}\sum_{i=1}^n  f_{\boldsymbol{\boldsymbol\theta}}^{\alpha}(Y_i),
\label{EQ:obj_func_MDPDE}
\end{eqnarray}
where $G_n$ is the empirical distribution function.
At $\alpha=0$, this MDPDE coincides with the MLE; 
the MDPDEs become more robust but less efficient as $\alpha$ increases, 
although the extent of loss is not significant in most cases with small $\alpha>0$;   
see \cite{Basu/etc:1998} for more details.

Under the censored data set-up of Section \ref{SEC:intro}, 
let us model the true distribution $G_X$ by the parametric model family 
$\mathcal{F}=\{F_{\boldsymbol{\boldsymbol\theta}} : {\boldsymbol\theta}\in \Theta\subseteq\mathbb{R}^p\}$
and denote the density of $F_{\boldsymbol{\boldsymbol\theta}}$ by $f_{\boldsymbol{\boldsymbol\theta}}$.
Based on $n$ i.i.d.~randomly censored observations $\{Z_i, \delta_i\}_{i=1,\ldots,n}$, 
as suggested in \cite{Basu/etc:2006}, we define the MDPDE by using the KMPL estimator $\widehat{G_X}$ 
in place of $G_n$ in (\ref{EQ:obj_func_MDPDE}), so that the estimating equation is given by 
\begin{eqnarray}
\int \boldsymbol{u}_{\boldsymbol{\boldsymbol\theta}}(y) f_{\boldsymbol{\boldsymbol\theta}}^{1+\alpha}(y)dy 
-  \int  \boldsymbol{u}_{\boldsymbol{\boldsymbol\theta}}(y)f_{\boldsymbol{\boldsymbol\theta}}^{\alpha}(y)d\widehat{G_X}(y) 
=\boldsymbol{0},
\label{EQ:Est_eqn_MDPDE}
\end{eqnarray}
where $\boldsymbol{u}_{\boldsymbol{\boldsymbol\theta}}=\nabla \ln f_{\boldsymbol{\boldsymbol\theta}}$ 
is the likelihood score function, with $\nabla$ representing the gradient with respect to $\boldsymbol{\theta}$. 
It is clearly an M-estimator with a model dependent $\boldsymbol{\psi}$-function given by
\begin{eqnarray}
\boldsymbol\psi(x;{\boldsymbol\theta})= \boldsymbol\psi_\alpha(x;{\boldsymbol\theta})
=\int \boldsymbol{u}_{\boldsymbol{\boldsymbol\theta}}(y) f_{\boldsymbol{\boldsymbol\theta}}^{1+\alpha}(y)dy 
-  \boldsymbol{u}_{\boldsymbol{\boldsymbol\theta}}(x)f_{\boldsymbol{\boldsymbol\theta}}^{\alpha}(x).
\label{EQ:psi_MDPDE}
\end{eqnarray}

The MDPDE is also Fisher consistent; its estimating equation (\ref{EQ:Est_eqn_MDPDE}) is unbiased at the model. 
Further, unlike general M-estimators defined only through an estimating equation, 
the MDPDEs have a solution to the multiple root issue, since there is a proper underlying objective function.
However, for censored data, the MDPDE at $\alpha=0$ is, in a strict sense, 
not exactly the MLE as studied in \cite{Borgan:1984}, since we use the KMPL in place of the empirical distribution function.
But it is closely related to the estimator studied by \cite{Oakes:1986} who called it the ``approximate MLE (AMLE)'';
this AMLE will be our standard of comparison.
The consistency of the MDPDE has been proved \citep[][Theorem 3.1]{Basu/etc:2006} under much simpler Assumption (B1)--(B5) of Appendix \ref{APP:Assumptions}; 
its asymptotic normality follows directly from the general results of M-estimator.

\subsection{Robustness: Influence Function of the Estimators}
\label{SEC:IF_MDPDE}

The influence function (IF) is the most popular and classical tool for the theoretical assessment of robustness \citep{Hampel/etc:1986}. 
It indicates a (first order) approximation to the bias in the estimator caused by 
infinitesimal contamination at an outlying data point (contamination point)
and hence measures the stability of the estimators.
However, the IF of M-estimators and MDPDEs under random censoring has not been studied in detail, except for few expressions in \cite{Wang:1999}.
Here, we fill this gap up by developing the general IF theory of M-estimators  along with illustration for the MDPDEs.

Suppose $\boldsymbol{T}_{\boldsymbol\psi}(G_X) (= \boldsymbol{\theta}_0)$ denotes the statistical functional 
corresponding to the M-estimator with a given $\boldsymbol\psi$-function
at $G_X$, defined as a solution of (\ref{EQ:est-eqn-psiX}).
Consider the contaminated distribution $G_\epsilon=(1-\epsilon)G_X + \epsilon\wedge_{t}$,
where $\epsilon$ is the contamination proportion and $\wedge_{t}$ denotes the degenerate distribution 
at the contamination point $t$. 
Then,  substituting $\boldsymbol{T}_{\boldsymbol\psi}(G_\epsilon)$ for ${\boldsymbol\theta}_0$ and $G_\epsilon$ for $G_X$ in (\ref{EQ:est-eqn-psiX})
and differentiating with respect to $\epsilon$ at $\epsilon=0$, we get the IF of $\boldsymbol{T}_{\boldsymbol\psi}$ at $G_X$ to have the form
\begin{eqnarray}
\mathcal{IF}(t;\boldsymbol{T}_{\boldsymbol\psi}, G_X) 
= \left.\frac{\partial}{\partial\epsilon}\boldsymbol{T}_{\boldsymbol\psi}(G_\epsilon)\right|_{\epsilon=0} 
= \boldsymbol\Lambda(\boldsymbol\psi; \boldsymbol{T}_{\boldsymbol\psi}(G_X))^{-1}\boldsymbol\psi(t;\boldsymbol{T}_{\boldsymbol\psi}(G_X)).
\label{EQ:IF_Mest}
\end{eqnarray}
Clearly, M-estimators with bounded $\boldsymbol\psi$-functions have bounded IFs and hence are robust with respect to infinitesimal contaminations.
But, if $\boldsymbol\psi$ is unbounded,  the resulting M-estimator has unbounded IF, implying its non-robust nature.

Now, using the $\boldsymbol{\psi}$-function from (\ref{EQ:psi_MDPDE}) in Equation (\ref{EQ:IF_Mest}), 
we get the IF of the MDPDE functional $\boldsymbol{T}_{\boldsymbol\psi_\alpha}$ 
which has the simple form at the model $G_X=F_{{\boldsymbol\theta}_0}$ as given by
\begin{eqnarray}
\mathcal{IF}(t;\boldsymbol{T}_{\boldsymbol\psi_\alpha}, F_{{\boldsymbol\theta}_0}) 
&=& \left(\int \boldsymbol{u}_{{\boldsymbol\theta}_0}(y) \boldsymbol{u}_{{\boldsymbol\theta}_0}^T(y) 
f_{{\boldsymbol\theta}_0}^{1+\alpha}(y)dy\right)^{-1}
\left[\int \boldsymbol{u}_{{\boldsymbol\theta}_0}(y) f_{{\boldsymbol\theta}_0}^{1+\alpha}(y)dy 
-  \boldsymbol{u}_{{\boldsymbol\theta}_0}(t)f_{{\boldsymbol\theta}_0}^{\alpha}(t)\right].\nonumber
\label{EQ:IF_MDPDE}
\end{eqnarray}
Note that, the IF of the MDPDE with $\alpha>0$ is bounded for most parametric models,
whereas it is unbounded at $\alpha=0$ (non-robust AMLE).
Hence, the MDPDEs with $\alpha>0$ yield robust estimators.

\begin{figure}[h!t]
	\centering
	\subfloat[IF]{
		\includegraphics[width=0.4\textwidth]{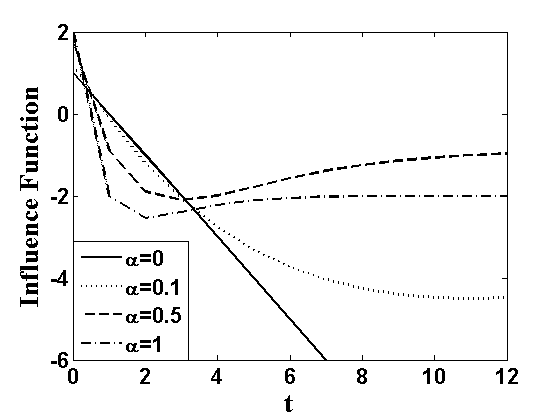}
		\label{FIG:IF_MDPDE_theta1}}
	~ 
	\subfloat[Empirical MSE]{
		\includegraphics[width=0.4\textwidth]{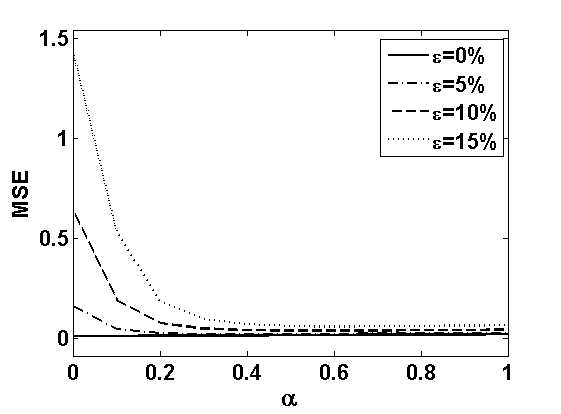}
		\label{FIG:MSE_MDPDE_theta1}}
	\caption{IFs and empirical MSE of the MDPDEs under exponential model (${\theta}_0=1$) with Exp(9) censoring.
	The $100\epsilon\%$ contamination is taken from Exp(10) for MSE calculation.}
	\label{FIG:MDPDE_exp}
\end{figure}

\bigskip\noindent
\textbf{Example \ref{SEC:IF_MDPDE}.1 [MDPDE under the Exponential Model].}
Consider the exponential model family (Exp($\theta$)) having mean ${\theta}>0$ (parameter of interest)	
and distribution function $F_{\theta}(x) = 1 - e^{-\frac{x}{{\theta}}}$ at $x>0$. 
In this case, the $\boldsymbol{\psi}$-function corresponding to the MDPDE with tuning parameter $\alpha>0$ is given by 
$
\psi_\alpha(x;{\theta}) = \frac{({\theta}-x)}{{\theta}^{\alpha+2}}e^{-\frac{\alpha x}{{\theta}}} 
- \frac{\alpha}{(1+\alpha)^2{\theta}^{\alpha+1}},
$
and hence 
$
\Lambda(\psi_\alpha; {\theta}_0)= {(1+\alpha^2)}{(1+\alpha)^{-3}}{\theta}^{-(\alpha+2)}.
$
Therefore, the IF of the resulting MDPDE ($T_{\psi_\alpha}$) at $G_X=F_{{\theta}_0}$ is given by 
\begin{eqnarray}
\mathcal{IF}(t;T_{\psi_\alpha}, F_{{\theta}_0}) &=& \frac{(1+\alpha)^3}{(1+\alpha^2)}
\left[({\theta}_0-t)e^{-\frac{\alpha t}{{\theta}_0}} - \frac{\alpha{\theta}_0}{(1+\alpha)^2}\right].\nonumber
\end{eqnarray}
Note that, these IFs are bounded for all $\alpha>0$ implying the robust nature of those MDPDEs.
However, at $\alpha=0$, we get $\mathcal{IF}(t;T_{\psi_0}, F_{{\theta}_0}) = ({\theta}_0-t)$
which, being a linear function, is clearly unbounded and implies the non-robust nature of the classical AMLE. 
Figure \ref{FIG:IF_MDPDE_theta1} shows their plots with ${\theta}_0=1$; 
the clear descending nature of the IFs with increasing $\alpha$ further implies their increasing robustness strengths.

On the other hand, it has been empirically illustrated in \cite{Basu/etc:2006} that the pure-data efficiency of the MDPDE 
decreases slightly with increasing $\alpha>0$, but its robustness under contamination increases significantly 
yielding smaller finite sample MSE. In Figure \ref{FIG:MSE_MDPDE_theta1}, 
we also present the empirical MSE of the MDPDEs from a simulation exercise, with 1000 replications, 
under this Exp($\theta$) model with sample size $n=100$, true parameter $\theta_{0}=1$, 
10\% (expected) exponential censoring (requires censoring mean to be $9$) and different proportions of contamination from Exp($10$).
The trade-offs between the efficiency and robustness of the MDPDEs over the tuning parameter $\alpha$,
and its advantages under contaminated observations are clearly observed from these figures. 
\hfill{$\square$}

\section{Consistent Estimation of the Asymptotic Variance of M-Estimators}\label{SEC:Var_Est}

As discussed earlier, the major challenge in developing any test is to get a consistent estimate of the covariance matrix of the estimator to be used. 
So, we first need a consistent estimate $\widehat{\boldsymbol\Sigma}(\boldsymbol\psi; {\boldsymbol\theta})$ 
of the covariance matrix $\boldsymbol\Sigma(\boldsymbol\psi; {\boldsymbol\theta})$ of the robust M-estimator
for a given $\boldsymbol{\psi}$-function based on the censored observations $\{Z_i, \delta_i\}$;
here we develop it under the fully parametric model setting.

Under the notation of Section \ref{SEC:Est}, let $G_X = F_{{\boldsymbol\theta}_0}$
with ${\boldsymbol\theta}_0$ being the true parameter value and suppose Assumption (A3) holds;
then $\int \boldsymbol\psi(x;{\boldsymbol\theta}_0)dG_X(x) =  \int \boldsymbol\psi(x;{\boldsymbol\theta}_0)dF_{{\boldsymbol\theta}_0}(x)= 0$.
Note that, in view of Slutsky's theorem, 
it is enough to consistently estimate $\Lambda(\boldsymbol\psi;{\boldsymbol\theta}_0)$ 
and $\boldsymbol{C}(\boldsymbol\psi;{\boldsymbol\theta}_0)$ separately.

Now, under our parametric model assumption, one can easily derive a closed form expression of 
$\Lambda(\boldsymbol\psi;{\boldsymbol\theta}_0)
=\int \frac{\partial}{\partial{\boldsymbol\theta}} \boldsymbol\psi(x;{\boldsymbol\theta}_0)dF_{{\boldsymbol\theta}_0}(x)$.
So, assuming its continuity in ${\boldsymbol\theta}$ and consistency of the M-estimator $\widehat{{\boldsymbol\theta}}_n$, 
a consistent estimate of $\Lambda(\boldsymbol\psi;{\boldsymbol\theta}_0)$ is given by $\Lambda(\boldsymbol\psi;\widehat{{\boldsymbol\theta}}_n)$.
However, for M-estimators without the fully parametric model assumption, 
a non-parametric estimate of $\boldsymbol\Lambda(\boldsymbol\psi;{\boldsymbol\theta})$ can also be obtained as
$\widehat{\boldsymbol\Lambda}_n(\boldsymbol\psi;\widehat{{\boldsymbol\theta}}_n) 
= \frac{1}{n} \sum_{i=1}^n \frac{\partial}{\partial{\boldsymbol\theta}} \boldsymbol\psi(X_i;{\boldsymbol\theta})$.

Next, the  harder challenge is to estimate $\boldsymbol{C}(\boldsymbol\psi;{\boldsymbol\theta})$ 
which depends on the unknown censoring distribution $G_C$ through $G_Z$ and 
hence cannot be computed explicitly, as in the case of $\Lambda(\boldsymbol\psi;{\boldsymbol\theta})$. 
To estimate it, we consider its defining function $U_{Z,\delta}(\cdot)$ 
which is further defined in terms of the quantities $\gamma_0$, $\gamma_1$, $\gamma_2$ and $\gamma$
involving the  unknown distributions $G_Z$, $G_{Z,0}$ and $G_{Z,1}$. 
However, these distributions can be estimated empirically, respectively, as
$$
\widehat{G}_{Z,n}(z) = \frac{1}{n}\sum_{i=1}^n I(Z_i\leq z)
~~~\mbox{and }~~~\widehat{G}_{Z,j,n}(z) = \frac{1}{n}\sum_{i=1}^n I(Z_i\leq z, \delta_i = j),~~ j=0,1.
$$
Note that, $\widehat{G}_{Z,n}$ and $\widehat{G}_{Z,j,n}$  are uniformly consistent for $G_Z$ and $G_{Z,j}$ respectively for $j=0,1$.
Plugging $\widehat{G}_{Z,n}(z)$ and $\widehat{G}_{Z,j,n}(z)$ in the definitions of $\gamma_0$, $\gamma_1$, $\gamma_2$ and $\gamma$,
we get their consistent estimators which we denote, respectively, as $\widehat{\gamma}_{0,n}$, $\widehat{\gamma}_{1,n}$, 
$\widehat{\gamma}_{2,n}$ and $\widehat{\gamma}_{n}$. 
At the ordered observations $\{Z_{(i,n)},\delta_{[i,n]}\}$, $i=1,\ldots,n$, they have the explicit forms given by 
\begin{eqnarray}
\widehat{\gamma}_{0,n}(Z_{(i,n)}) &=& \exp\left\{\sum_{j=1}^{i-1} \frac{I(\delta_{[j,n]}=0)}{n-j}\right\},
~~~\widehat{\gamma}_{n}(Z_{(i,n)}) = \sum_{j=1}^{i-1} \frac{n I(\delta_{[j,n]}=0)}{(n-j)^2},\nonumber\\
\widehat{\gamma}_{1,n}(Z_{(i,n)};\phi) &=& 	
\frac{1}{n-i+1}\sum_{j=i+1}^{n} I(\delta_{[j,n]}=1)\phi(Z_{(j,n)})\widehat{\gamma}_{0,n}(Z_{(j,n)}),\nonumber
\end{eqnarray}
\begin{eqnarray}
\widehat{\gamma}_{2,n}(Z_{(i,n)};\phi) &=& 	
\frac{1}{n}\left[\sum_{j=1}^{i} I(\delta_{[j,n]}=1)\widehat{\gamma}_{n}(Z_{(j,n)})\phi(Z_{(j,n)})\widehat{\gamma}_{0,n}(Z_{(j,n)})\right.
\nonumber\\
&& ~~~+ \left.\widehat{\gamma}_{n}(Z_{(i,n)})\sum_{j=i+1}^{n} I(\delta_{[j,n]}=1)\phi(Z_{(j,n)})\widehat{\gamma}_{0,n}(Z_{(j,n)})\right].
\nonumber
\end{eqnarray}
Then, assuming continuity of $\boldsymbol\psi$ in ${\boldsymbol\theta}$, 
a consistent estimate of the function $U_{Z,\delta}(\psi_j(\cdot, {\boldsymbol\theta}))$
is given by $\widehat{U}_{Z,\delta}(\psi_j(\cdot;\widehat{{\boldsymbol\theta}}_n))$ for each $j=1,\ldots,p$, where
$
\widehat{U}_{Z,\delta}(\phi)
=\phi(Z)\widehat{\gamma}_{0,n}(Z)\delta + \widehat{\gamma}_{1,n}(Z;\phi)(1-\delta) - \widehat{\gamma}_{2,n}(Z;\phi).
$
Thus, we finally get a consistent estimator of $\boldsymbol{C}(\boldsymbol\psi;{\boldsymbol\theta})$ as 
\begin{eqnarray}
\widehat{\boldsymbol{C}}_n(\boldsymbol\psi;\widehat{{\boldsymbol\theta}}_n) 
&=& \frac{1}{n}\sum_{i=1}^n \widehat{\boldsymbol{U}}(Z_i,\delta_i; \boldsymbol{\psi}(\cdot;\widehat{{\boldsymbol\theta}}_n))
\widehat{\boldsymbol{U}}(Z_i,\delta_i; \boldsymbol{\psi}(\cdot;\widehat{{\boldsymbol\theta}}_n))^T
\nonumber\\
&=& \frac{1}{n}\sum_{i=1}^n \widehat{\boldsymbol{U}}(Z_{(i,n)},\delta_{[i,n]}; \boldsymbol{\psi}(\cdot;\widehat{{\boldsymbol\theta}}_n))
\widehat{\boldsymbol{U}}(Z_{(i,n)},\delta_{[i,n]}; \boldsymbol{\psi}(\cdot;\widehat{{\boldsymbol\theta}}_n))^T
\label{EQ:C_psi_est}
\end{eqnarray}
where $\widehat{\boldsymbol{U}}(Z,\delta, \boldsymbol{\psi}(\cdot;\widehat{{\boldsymbol\theta}}_n))= 
\left(\widehat{U}_{Z,\delta}(\psi_1(\cdot;\widehat{{\boldsymbol\theta}}_n)), \cdots, \widehat{U}_{Z,\delta}(\psi_p(\cdot;\widehat{{\boldsymbol\theta}}_n))\right)^T$.

Therefore, the asymptotic covariance matrix $\boldsymbol\Sigma(\boldsymbol\psi; {\boldsymbol\theta}_0)$ 
of a consistent sequence of M-estimators $\widehat{{\boldsymbol\theta}}_n$ can be estimated consistently by
the estimator 
$
\widehat{\boldsymbol\Sigma}_n(\boldsymbol\psi; \widehat{{\boldsymbol\theta}}_n) 
= \boldsymbol\Lambda(\boldsymbol\psi;\widehat{{\boldsymbol\theta}}_n)^{-1}
\widehat{C}_n(\boldsymbol\psi;\widehat{{\boldsymbol\theta}}_n)\boldsymbol\Lambda(\boldsymbol\psi;\widehat{{\boldsymbol\theta}}_n)^{-1},
$
or the estimator
$
\widehat{\boldsymbol\Sigma}_n(\boldsymbol\psi; \widehat{{\boldsymbol\theta}}_n) 
= \widehat{\boldsymbol\Lambda}_n(\boldsymbol\psi;\widehat{{\boldsymbol\theta}}_n)^{-1}
\widehat{C}_n(\boldsymbol\psi;\widehat{{\boldsymbol\theta}}_n)
\widehat{\boldsymbol\Lambda}_n(\boldsymbol\psi;\widehat{{\boldsymbol\theta}}_n)^{-1}.
$

Let us now illustrate the performance and rate of consistency of the proposed variance estimator
for the MDPDE under the class of Weibull model.

\bigskip\noindent
\textbf{Example \ref{SEC:Var_Est}.1 [MDPDE under Weibull Model].}
Consider the Weibull model with parameter $\boldsymbol{{\boldsymbol{\theta}} }=(a, b)^T$
as described in the motivating examples in Section \ref{SEC:intro}.
The distribution function of this Weibull($a,b$) model is given by 
$F_{\boldsymbol{\theta}}(x) = 1 - e^{-(ax)^b}$ at $x>0$,
so that the MDPDE $\boldsymbol{\psi}$-function at an $\alpha>0$ is given by (\ref{EQ:psi_MDPDE})
with density $f_{\boldsymbol{\theta}}(x) = b a^b x^{b-1}e^{-(ax)^b}$ and score function 
$$
\boldsymbol{u}_{\boldsymbol{\theta}}(x) = \begin{bmatrix}
\begin{array}{cc}
\frac{b}{a}(1 - (ax)^b)\\
\ln(ax)(1 - (ax)^b) + \frac{1}{b}
\end{array}\end{bmatrix}.
$$
Note that the integral in (\ref{EQ:psi_MDPDE}) and hence the $\boldsymbol{\psi}$-function do not have a close form expression in this Weibull model,
unlike the exponential model in Example 2.3.1; they need to be computed numerically.
Similarly, the matrix $\boldsymbol{\Lambda}(\boldsymbol{\psi}_\alpha; \boldsymbol{\theta}_0)$ cannot be expressed in a closed form 
and so, after computation of the MDPDE, $\widehat{{\boldsymbol\theta}}_n=(\widehat{a}_n, \widehat{b}_n)^T$,
we can estimate it by $\widehat{\boldsymbol\Lambda}_n(\boldsymbol\psi;\widehat{{\boldsymbol\theta}}_n)$.
A consistent variance estimate of this MDPDE can then be obtained following our proposal described above. 

\begin{figure}[h!t]
	\centering
	\subfloat[Paramter $a$]{
		\includegraphics[width=0.45\textwidth]{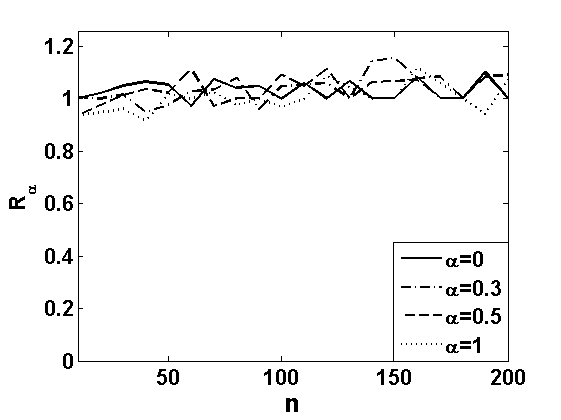}
		\label{FIG:Var_Est_Ratio_Par1}}
	~ 
	\subfloat[Paramter $b$]{
		\includegraphics[width=0.45\textwidth]{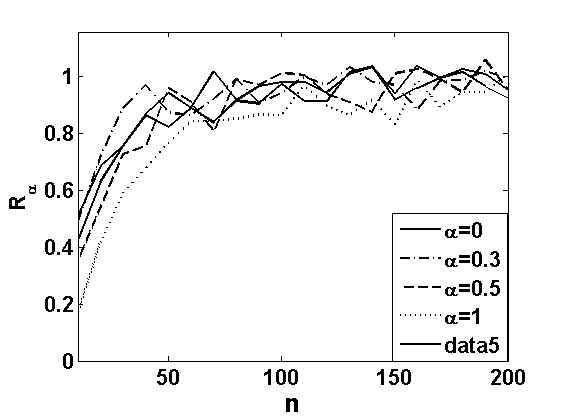}
		\label{FIG:Var_Est_Ratio_Par2}}
	\caption{The ratio ($R_\alpha$) of variance estimator and empirical MSE of the MDPDEs of $\boldsymbol{{\boldsymbol{\theta}} }=(a, b)^T$
		under Weibull($a, b$) model with true parameters $(a,b)=(2,5)$ and $10\%$ censoring from Exp(17.4). }
	\label{FIG:MDPDE_Weib}
\end{figure}

Let us compute the MDPDE and its variance estimate based on simulated samples of various sizes ($n$) drawn from Weibull(2,5) distribution 
along with (expected) 10\% censoring from exponential distribution (this requires the censoring mean to be around 17.4).
The process is repeated 1000 times to compute the empirical MSE of the MDPDEs and the average of their variance estimates; 
their ratio (say, $R_\alpha$) is then plotted over $n$ for different $\alpha$ in Figure \ref{FIG:MDPDE_Weib}. 
Note that, a value of $R_\alpha$ close to one indicates the consistency of our proposed variance estimator for the MDPDE
which is seen to hold for a reasonably moderate sample size $n=200$.
This ratio is, in fact, quite close to one for all $n\geq 10$ for the scale parameter $a$ 
whereas we need slightly larger sample sizes $n > 50$ for the shape parameter $b$ in the Weibull model.
These observations clearly indicate the desired performance of the proposed variance estimator
to be useful in developing robust tests of hypotheses in subsequent sections.
\hfill{$\square$}

\section{Robust Tests for One-Sample Problems}\label{SEC:One_test}

\subsection{The Wald-Type Tests Statistics}\label{SEC:Wald_test}

Consider the set-up of randomly censored observations $\{Z_i, \delta_i\}_{i=1,\ldots,n}$ as in the previous sections with  $G_X\in\mathcal{F}$.
The simplest possible  hypothesis is then given by
\begin{equation}
H_0:~\boldsymbol{{\boldsymbol\theta}}=\boldsymbol{{\boldsymbol\theta}}_0, ~~~~\mbox{ against }~~~~ H_1:~\boldsymbol{{\boldsymbol\theta}} \neq \boldsymbol{{\boldsymbol\theta}}_0,
\label{EQ:Hyp_simple}
\end{equation}
where $\boldsymbol{{\boldsymbol\theta}}_0\in \Theta$ is prefixed.
However, we here consider a more general class of parametric hypothesis, also containing this simple one in (\ref{EQ:Hyp_simple}), 
as given by 
\begin{equation}
H_{0}:\boldsymbol{{\boldsymbol\theta}}\in \Theta_{0},~~~~\text{ against }~~ 
H_{1}:\boldsymbol{{\boldsymbol\theta}}\in \Theta-\Theta_{0},
\label{EQ:Hyp_comp}
\end{equation}
where  $\Theta_{0}$ is a fixed proper subset of $\Theta$.
In most applications, $\Theta_{0}$ is defined in terms of $r(\leq p)$ restrictions of the form
$\boldsymbol{m}(\boldsymbol{{\boldsymbol\theta}})=\boldsymbol{0}$
for some function $\boldsymbol{m}:\mathbb{R}^{p}\rightarrow \mathbb{R}^{r}$.
We assume that the $p\times r$ matrix
$\boldsymbol{M}\left(  \boldsymbol{{\boldsymbol\theta}}\right)  
=\frac{\partial\boldsymbol{m}(\boldsymbol{{\boldsymbol\theta}})}{\partial \boldsymbol{{\boldsymbol\theta}}}$
exists and is continuous in $\boldsymbol{{\boldsymbol\theta}}$ with 
rank$\left(\boldsymbol{M}\left(  \boldsymbol{{\boldsymbol\theta}}\right)\right)  =r.$ 
Then, we construct the Wald-type test statistics for testing (\ref{EQ:Hyp_comp}),
based on the robust M-estimators $\widehat{\boldsymbol{{\boldsymbol\theta}}}_n$ of $\boldsymbol{{\boldsymbol\theta}}$
associated with a given $\boldsymbol{\psi}$-function, as given by
\begin{equation}
W_{n}^\ast = n \boldsymbol{m}^{T}\left(\widehat{\boldsymbol{{\boldsymbol\theta}}}_{n}\right)  
\left[  \boldsymbol{M}^{T}(\widehat{\boldsymbol{{\boldsymbol\theta}}}_{n})
\widehat{\boldsymbol{\boldsymbol\Sigma}}_{n}(\boldsymbol\psi;\widehat{\boldsymbol{{\boldsymbol\theta}}}_{n})
\boldsymbol{M}(\widehat{\boldsymbol{{\boldsymbol\theta}}}_{n})\right]^{-1}
\boldsymbol{m}\left(\widehat{\boldsymbol{{\boldsymbol\theta}}}_{n}\right),
\label{EQ:Wald_TS_comp}
\end{equation}
where $\widehat{\boldsymbol{\boldsymbol\Sigma}}_n(\boldsymbol\psi; \widehat{\boldsymbol{{\boldsymbol\theta}}}_n)$ 
is a consistent estimator of the variance of  $\widehat{\boldsymbol{{\boldsymbol\theta}}}_n$  from Section \ref{SEC:Var_Est}.

Note that, the simple hypothesis given in (\ref{EQ:Hyp_simple}) is a special case of (\ref{EQ:Hyp_comp})
with $\Theta_0 = \{\boldsymbol{\theta}_0\}$, $r=p$, $\boldsymbol{m}(\boldsymbol{\theta}) = \boldsymbol{\theta} - \boldsymbol{\theta}_0$
and $\boldsymbol{M}(\boldsymbol{\theta}) = \boldsymbol{I}_p$, the identity matrix of order $p$;
then $W_n^\ast$ has a much simpler form as 
$W_n^\ast = W_{n}^0 = n(\widehat{\boldsymbol{{\boldsymbol\theta}}}_{n}-\boldsymbol{{\boldsymbol\theta}}_{0})^{T}
\widehat{\boldsymbol{\boldsymbol\Sigma}}_n(\boldsymbol\psi; \widehat{\boldsymbol{{\boldsymbol\theta}}}_n)^{-1}
(\widehat{\boldsymbol{{\boldsymbol\theta}}}_{n}-\boldsymbol{{\boldsymbol\theta}}_{0}).$ 

Now, using consistency of $\widehat{\boldsymbol{\boldsymbol\Sigma}}_n(\boldsymbol\psi; \widehat{\boldsymbol{{\boldsymbol\theta}}}_n)$,
one can show along the lines of \cite{Basu/etc:2016} that, $W_{n}^\ast$ asymptotically follows a chi-square distribution ($\chi_r^2$)
with $r$ degrees of freedom under the null hypothesis in (\ref{EQ:Hyp_comp}).
Further, a power approximation at $\alpha_0$-significance level can be obtained as given by 

$$\pi_{W_{n}^\ast}\left(\boldsymbol{{\boldsymbol\theta}}^{\ast}\right)
=P \left(  W_{n}^\ast>\chi_{r,\alpha}^{2}|\boldsymbol{{\boldsymbol\theta}}=\boldsymbol{{\boldsymbol\theta}}^{\ast}\right)
\approx 1-\Phi\left(\frac{n^{1/2}}{\sigma_{\ast}\left(\boldsymbol{{\boldsymbol\theta}}^{\ast}\right)}
\left(  \frac{\chi_{r,\alpha}^{2}}{n} - \bar{W}\left(\boldsymbol{{\boldsymbol\theta}}^{\ast}\right)\right)\right),
~\mbox{ for any } \boldsymbol{{\boldsymbol{\theta}} }^\ast\notin \Theta_0, 
$$
where $\Phi(\cdot)$ is the standard normal distribution function,
$\chi_{r,\alpha_0}^{2}$ denote the $(1-\alpha_0)$-th quantile of $\chi_r^2$,
$\bar{W}\left(\boldsymbol{{\boldsymbol\theta}}\right)  
=\boldsymbol{m}^{T}(\boldsymbol{{\boldsymbol\theta}})\left[\boldsymbol{M}^{T}(\boldsymbol{{\boldsymbol\theta}})
\boldsymbol{\boldsymbol\Sigma}(\boldsymbol\psi;\boldsymbol{{\boldsymbol\theta}})
\boldsymbol{M}(\boldsymbol{{\boldsymbol\theta}})\right]^{-1}\boldsymbol{m}(\boldsymbol{{\boldsymbol\theta}})$,
and
$
\sigma_{\ast}^{2}\left(\boldsymbol{{\boldsymbol\theta}}^{\ast}\right)  
=\left(\frac{\partial \bar{W}\left(\boldsymbol{{\boldsymbol\theta}}\right)}{\partial\boldsymbol{{\boldsymbol\theta}}}
\right)_{\boldsymbol{{\boldsymbol\theta}={\boldsymbol\theta}}^{\ast}}^{T}
\boldsymbol{\boldsymbol\Sigma}(\boldsymbol\psi;\boldsymbol{{\boldsymbol\theta}}^{\ast})
\left(\frac{\partial \bar{W}\left(\boldsymbol{{\boldsymbol\theta}}\right)}{\partial\boldsymbol{{\boldsymbol\theta}}}
\right)_{\boldsymbol{{\boldsymbol\theta}={\boldsymbol\theta}}^{\ast}}.
$
This approximation can be used to derive the sample size requirement while designing the clinical trial 
to achieve a desired power in their robust analysis. 
Also, for any $\boldsymbol{{\boldsymbol\theta}}^{\ast}\notin \Theta_0$, 
$\lim\limits_{n\rightarrow \infty}\pi_{W_{n}^\ast}\left(\boldsymbol{{\boldsymbol\theta}}^{\ast}\right)=1$,
implying the consistency of the proposed Wald-type tests.

Again, in the line of \cite{Basu/etc:2016}, one can show that the asymptotic distribution
of $W_n^\ast$ under the contiguous alternative hypotheses 
$H_{1,n}:\boldsymbol{\theta} = \boldsymbol{{\boldsymbol\theta}}_{n}$, where $\boldsymbol{{\boldsymbol\theta}}_{n}=\boldsymbol{{\boldsymbol\theta}}_{0}+n^{-1/2}\boldsymbol{d}$,
with $\boldsymbol{d}\in \mathbb{R}^{p} - \{\mathbf{0}_p\}$,
is $\chi_{r}^{2}\left(\delta\right)$, the non-central chi-square distribution with $r$ degrees of freedom and non-centrality parameter
$\delta$, with 
$\delta=\boldsymbol{d}^{T}\boldsymbol{M}\left(\boldsymbol{{\boldsymbol\theta}}_{0}\right)
\boldsymbol{\Sigma}^\ast(\boldsymbol\psi;\boldsymbol{\theta}_{0})^{-1}\boldsymbol{M}
\left(\boldsymbol{{\boldsymbol\theta}}_{0}\right)^{T}\boldsymbol{d}$,
where $\boldsymbol{\Sigma}^\ast(\boldsymbol\psi;\boldsymbol{\theta})=\boldsymbol{M}^{T}(\boldsymbol{{\boldsymbol\theta}})
\boldsymbol{\boldsymbol\Sigma}(\boldsymbol\psi;\boldsymbol{\theta})\boldsymbol{M}(\boldsymbol{{\boldsymbol\theta}})$.
This will help us to directly obtain the asymptotic contiguous power of our proposed tests for known censoring  
or estimate it for unknown censoring mechanism using the consistent estimators of $\boldsymbol{\theta}$ and 
$\boldsymbol{\boldsymbol\Sigma}(\boldsymbol\psi;\boldsymbol{\theta})$.

\smallskip\noindent
\textbf{Example \ref{SEC:Wald_test}.1 [MDPDE based Tests under Weibull Model].}
Consider the simulation study as in Example \ref{SEC:Var_Est}.1 for the MDPDE under Weibull($a, b$) model
with 10\% exponential censoring. Let us now perform the proposed MDPDE based test 
for two simple hypotheses $H_0^{(1)}: (a, b) = (2, 5)$, $H_0^{(2)}: (a, b) = (2.2, 2.3)$
and two composite hypotheses $H_0^{(3)}: b=5$, $H_0^{(4)}: b=2$, assuming unknown $a$, against their respective omnibus alternatives
at the 5\% level of significance. We present the proportion of rejection among 1000 replications,
at sample size $n=100$, in Table \ref{TAB:Test_Weib}; note that the hypotheses $H_0^{(1)}$ and $H_0^{(3)}$
leads to level of the tests whereas the other two hypotheses give the power against contiguous alternatives. 
Additionally, to study the robustness of the proposed tests, we repeat the above simulation exercise by contaminating 5\% 
of each samples by observations from exponential distribution with mean $5$;
the corresponding level and power under contamination are also presented in Table \ref{TAB:Test_Weib}.
Clearly, even at a slight contamination of $5\%$, the level and power of the AMLE based classical Wald test (at $\alpha=0$)
essentially break down and differ drastically from their desired values, 
whereas the proposed MDPDE based Wald-type tests produce highly stable level and power for $\alpha \geq 0.3$.
\hfill{$\square$}

\begin{table}[h]
\centering
\caption{Simulated proportion of Rejection (level or power) by the MDPDE based Wald-type tests
		for different hypotheses under Weibull model with exponential censoring ($n=100$)}
\begin{tabular}{r| cccc | cccc} \hline
$\alpha$		& \multicolumn{4}{c}{Pure Data} & 		\multicolumn{4}{c}{5\% contaminated Data}\\
&$H_0^{(1)}$ & $H_0^{(2)}$  &	$H_0^{(3)}$ 	&	$H_0^{(4)}$ &$H_0^{(1)}$ & $H_0^{(2)}$  &	$H_0^{(3)}$ 	&	$H_0^{(4)}$ 	\\\hline
0	&	0.070	&	1	&	0.049	&	1	&	0.901	&	0.313	&	0.917	&	0.247	\\
0.1	&	0.055	&	1	&	0.038	&	1	&	0.150	&	0.988	&	0.194	&	0.992	\\
0.2	&	0.053	&	1	&	0.034	&	1	&	0.096	&	1	&	0.108	&	1	\\
0.3	&	0.054	&	1	&	0.032	&	1	&	0.087	&	1	&	0.085	&	1	\\
0.4	&	0.056	&	1	&	0.033	&	1	&	0.082	&	1	&	0.075	&	1	\\
0.5	&	0.058	&	1	&	0.038	&	1	&	0.074	&	1	&	0.075	&	1	\\
0.6	&	0.061	&	1	&	0.037	&	1	&	0.072	&	1	&	0.071	&	1	\\
0.7	&	0.061	&	1	&	0.041	&	1	&	0.070	&	1	&	0.070	&	1	\\
0.8	&	0.059	&	1	&	0.043	&	1	&	0.069	&	1	&	0.069	&	1	\\
0.9	&	0.061	&	1	&	0.043	&	1	&	0.073	&	1	&	0.072	&	1	\\
1	&	0.063	&	1	&	0.037	&	1	&	0.069	&	1	&	0.072	&	1	\\
		\hline
	\end{tabular}
	\label{TAB:Test_Weib}
\end{table}

\subsection{Robustness: Influence Analysis}\label{SEC:IF_test}

In order to study the robustness of $W_n^\ast$ through its IF \citep{Hampel/etc:1986},
we define the corresponding statistical functional as (ignoring the multiplier $n$)%
\begin{equation}
W_{\boldsymbol\psi}^\ast(G_X)=\boldsymbol{m}^{T}(\boldsymbol{T}_{\boldsymbol\psi}(G_X))
\boldsymbol{\Sigma}^\ast(\boldsymbol\psi;\boldsymbol{T}_{\boldsymbol\psi}(G_X))^{-1}\boldsymbol{m}(\boldsymbol{T}_{\boldsymbol\psi}(G_X)).
\label{WF}%
\end{equation}
Let $\boldsymbol{{\boldsymbol\theta}}_{0}\in\Theta_{0}$ be the true parameter value under (\ref{EQ:Hyp_comp}); 
we have $G_X=F_{\boldsymbol{{\boldsymbol\theta}}_{0}}$ and 
$\boldsymbol{m}(\boldsymbol{T}_{\boldsymbol\psi}(G_X))=\boldsymbol{0}_{r}$ for all $\boldsymbol\psi$-functions under consideration.
Hence, the first order IF of $W_{\boldsymbol\psi}^\ast(\cdot)$ at the null hypothesis in (\ref{EQ:Hyp_comp}) turns out to be 
$
\mathcal{IF}(t,W_{\boldsymbol\psi}^\ast,F_{\boldsymbol{{\boldsymbol\theta}}_{0}})=0.
$
So, we need to consider the second order IF of $W_{\boldsymbol\psi}^\ast$ at the null hypothesis in (\ref{EQ:Hyp_comp}), 
which is given  by
	\begin{eqnarray}
&&\mathcal{IF}_2({t},W_{\boldsymbol\psi}^{\ast},F_{\boldsymbol{{\boldsymbol\theta}}_{0}})
= 2\mathcal{IF}(t,\boldsymbol{T}_{\boldsymbol\psi},F_{\boldsymbol{{\boldsymbol\theta}}_{0}})^{T}
\boldsymbol{M}(\boldsymbol{{\boldsymbol\theta}}_{0})
\boldsymbol{\boldsymbol\Sigma}^\ast(\boldsymbol\psi;\boldsymbol{{\boldsymbol\theta}}_{0})^{-1}
\boldsymbol{M}^{T}(\boldsymbol{{\boldsymbol\theta}}_{0})
\mathcal{IF}(t,\boldsymbol{T}_{\boldsymbol\psi},F_{\boldsymbol{{\boldsymbol\theta}}_{0}})\nonumber\\
&& ~~~~~= 2\boldsymbol\psi(t;{\boldsymbol\theta}_0)^{T}
\boldsymbol\Lambda(\boldsymbol\psi; {\boldsymbol\theta}_0)^{-1}\boldsymbol{M}(\boldsymbol{{\boldsymbol\theta}}_{0})
\boldsymbol{\boldsymbol\Sigma}^\ast(\boldsymbol\psi;\boldsymbol{{\boldsymbol\theta}}_{0})^{-1}
\boldsymbol{M}^{T}(\boldsymbol{{\boldsymbol\theta}}_{0})
\boldsymbol\Lambda(\boldsymbol\psi; {\boldsymbol\theta}_0)^{-1}\boldsymbol\psi(t;{\boldsymbol\theta}_0).
\label{EQ:IF_comp}
\end{eqnarray}

The above second order IF of $W_{\boldsymbol\psi}^\ast$ is bounded implying the robustness of the proposed Wald-type tests,
whenever the underlying $\boldsymbol\psi$-function is bounded.
But, it is non-robust for unbounded $\boldsymbol\psi$-functions.
In particular, the IFs  of the MDPDE based Wald-type tests can be derived 
as a special case; the corresponding second order IF is bounded for all $\alpha>0$ 
and unbounded for $\alpha=0$ in most parametric models. 
This proves the desired robust nature of these tests at $\alpha>0$ 
and the non-robust nature of the classical AMLE based Wald test at $\alpha=0$.

Let us now study the stability of power and size of the proposed Wald-type tests 
through the corresponding influence function analysis.
Since these tests are consistent,  we consider the asymptotic power under the contiguous alternatives $H_{1,n}$.
However, in order to derive the power and level influence functions, 
we also consider additional contiguous contamination, assuming $G_X = F_{\boldsymbol{{\boldsymbol\theta}}_0}$,
as
$F_{n,\epsilon,t}^{P}=\left(1-\frac{\epsilon}{\sqrt{n}}\right)F_{\boldsymbol{{\boldsymbol\theta}}_{n}}
+\frac{\epsilon}{\sqrt{n}}\wedge_t$, 
and
$F_{n,\epsilon,t}^{L}=\left(1-\frac{\epsilon}{\sqrt{n}}\right)F_{\boldsymbol{{\boldsymbol\theta}}_{0}}
+\frac{\epsilon}{\sqrt{n}}\wedge_t$, 
respectively \citep{Hampel/etc:1986}.
Let us denote the asymptotic level and power of the proposed Wald-type test statistics $W_n^\ast$ 
under these contaminated distributions as 
$\alpha_{W_{n}^\ast}(\epsilon,t)=\lim\limits_{n\rightarrow\infty}P_{F_{n,\epsilon,t}^{L}}(W_{n}^\ast>\chi_{r,\alpha}^{2})$
and
$\beta_{W_{n}^\ast}(\boldsymbol{{\boldsymbol\theta}}_{n},\epsilon,t)=\lim\limits_{n\rightarrow\infty}P_{F_{n,\epsilon, t}^{P}}(W_{n}^\ast>\chi_{r,\alpha}^{2})$.
Then, the level influence function (LIF) and the power influence function (PIF) of $W_n^\ast$ are defined as
\[
\mathcal{LIF}(t;W_{\boldsymbol\psi}^\ast,F_{\boldsymbol{{\boldsymbol\theta}}_{0}})=\left.\dfrac{\partial}{\partial\epsilon}\alpha_{W_{n}^\ast}(\epsilon,t)\right\vert_{\epsilon=0}
~~\mbox{ and }~~
\mathcal{PIF}(t;W_{\boldsymbol\psi}^\ast,F_{\boldsymbol{{\boldsymbol\theta}}_{0}})=\left.\dfrac{\partial}{\partial\epsilon}\beta_{W_{n}^\ast}(\boldsymbol{{\boldsymbol\theta}}_{n},\epsilon,t)\right\vert _{\epsilon=0}.
\]

Now, following a similar calculation as in \cite{Basu/etc:2016}, we can show that,
under $F_{n,\epsilon,t}^{P}$, $W_{n}^\ast$ asymptotically follows 
$\chi_r^2(\delta^\ast)$ distribution, where 		
$\delta^\ast  = \widetilde{\boldsymbol{d}}_{\epsilon}^{T}(\boldsymbol{{\boldsymbol\theta}}_{0}) 
\boldsymbol{M}(\boldsymbol{{\boldsymbol\theta}}_{0})\boldsymbol{\boldsymbol\Sigma}^{\ast}(\boldsymbol\psi;\boldsymbol{{\boldsymbol\theta}}_{0})^{-1}
\boldsymbol{M}^{T}(\boldsymbol{{\boldsymbol\theta}}_{0})\widetilde{\boldsymbol{d}}_{\epsilon}(\boldsymbol{{\boldsymbol\theta}}_{0}),$
with 
$\widetilde{\boldsymbol{d}}_{\epsilon}(\boldsymbol{{\boldsymbol\theta}}_{0})
=\boldsymbol{d}+\epsilon\mathcal{IF}(t,\boldsymbol{T}_{\boldsymbol\psi},F_{\boldsymbol{{\boldsymbol\theta}}_{0}})$. 
Then, the asymptotic power of $W_n^\ast$ under $F_{n,\epsilon,t}^{P}$ can be approximated as
		\begin{align}
		\beta_{W_{n}^{\ast}}(\boldsymbol{{\boldsymbol\theta}}_{n},\epsilon,t)  
		&  =\sum\limits_{v=0}^{\infty}C_{v}\left(  
		\boldsymbol{M}^{T}(\boldsymbol{{\boldsymbol\theta}}_{0})
		\widetilde{\boldsymbol{d}}_{\epsilon}(\boldsymbol{{\boldsymbol\theta}}_{0}),
		\boldsymbol{\boldsymbol\Sigma}^{\ast}(\boldsymbol\psi;\boldsymbol{{\boldsymbol\theta}}_{0})^{-1}\right)  
		P\left(  \chi_{r+2v}^{2}>\chi_{r,\alpha}^{2}\right),\nonumber
		\end{align}
where $C_{v}\left(  \boldsymbol{t},\boldsymbol{A}\right)  
=\frac{\left(\boldsymbol{t}^{T}\boldsymbol{A}\boldsymbol{t}\right)^{v}}{v!2^{v}}
e^{-\frac{1}{2}\boldsymbol{t}^{T}\boldsymbol{A}\boldsymbol{t}}.
$	
%
%
%
In particular, substituting $\boldsymbol{d}=\boldsymbol{0}_{p}$
we get the asymptotic level of $W_n^\ast$ under the contaminated distribution $F_{n,\epsilon,t}^{L}$ as given by
	$\alpha_{W_{n}^\ast}(\epsilon,t)  
	=\beta_{W_{n}^\ast}(\boldsymbol{{\boldsymbol\theta}}_{0},\epsilon,t) 
	=\sum\limits_{v=0}^{\infty}C_{v}\left(  \epsilon\boldsymbol{M}^{T}(\boldsymbol{{\boldsymbol\theta}}_{0})
	\mathcal{IF}(t,T_{\boldsymbol\psi},F_{\boldsymbol{{\boldsymbol\theta}}_{0}}),
	\boldsymbol{\boldsymbol\Sigma}^{\ast}(\boldsymbol\psi;\boldsymbol{{\boldsymbol\theta}}_{0})^{-1}\right)  
	P\left(\chi_{r+2v}^{2} > \chi_{r,\alpha}^{2}\right)$.

Now we can derive the LIF and PIF of $W_n^\ast$ 
by differentiating $\beta_{W_{n}^{\ast}}(\boldsymbol{{\boldsymbol\theta}}_{n},\epsilon,t)$ 
and $\alpha_{W_{n}^{\ast}}(\epsilon,t)$ with respect to $\epsilon$ at $\epsilon=0$, respectively.
Whenever $\mathcal{IF}(t,\boldsymbol{T}_{\boldsymbol\psi},F_{\boldsymbol{{\boldsymbol\theta}}_{0}})$ finite,
they are given by 
$$	\mathcal{LIF}(t,W_{\boldsymbol\psi}^\ast,F_{\boldsymbol{{\boldsymbol\theta}}_{0}})=0,
~~~ \mathcal{PIF}(t,W_{\boldsymbol\psi}^\ast,F_{\boldsymbol{{\boldsymbol\theta}}_{0}})
= K_{r}^{\ast}\left( \boldsymbol{S}_{0}\boldsymbol{d}\right)
\boldsymbol{S}_{0}\mathcal{IF}(t,\boldsymbol{T}_{\boldsymbol\psi},F_{\boldsymbol{{\boldsymbol\theta}}_{0}}),
$$
with $K_{p}^{\ast}(s)=e^{-\frac{s}{2}}\sum_{v=0}^{\infty}s^{v-1}2^{-v}
\left(  2v-s\right)  P\left(  \chi_{p+2v}^{2}>\chi_{p,\alpha}^{2}\right)/v!$
and $\boldsymbol{S}_{0} = \boldsymbol{d}^{T}\boldsymbol{M}(\boldsymbol{{\boldsymbol\theta}}_{0})
\boldsymbol{\boldsymbol\Sigma}^{\ast}(\boldsymbol\psi;\boldsymbol{{\boldsymbol\theta}}_{0})^{-1}
\boldsymbol{M}^{T}(\boldsymbol{{\boldsymbol\theta}}_{0}).$
Note  that the PIF of the proposed test is bounded implying the power stability under contiguous contamination, 
whenever the IF of the M-estimator used is bounded, 
i.e., whenever the underlying $\boldsymbol\psi$-function is bounded, and vice versa. 
In particular, the MDPDE based tests have stable asymptotic power with bounded PIF for all $\alpha>0$; 
the PIF of the classical AMLE based Wald test (at $\alpha=0$) is unbounded implying its non-robust nature.
Further, for bounded $\boldsymbol{\psi}$-functions, 
the LIF is identically zero which implies that the level of the corresponding Wald-type tests 
remain  asymptotically stable under a contiguous contamination.

\section{Further Useful Extensions}\label{SEC:Wald_test_Ext}
\subsection{Robust Comparison of Two Independent Censored Groups}\label{SEC:Wald_test2}

In any clinical trial study, we generally have two or more independent groups of patients 
exposed to different treatments  (and control) which need to be compared statistically.
Extending the idea from the previous section, we can also develop the Wald-type tests
for robust comparison of two such censored groups using the M-estimators (and MDPDE).
The corresponding uncensored cases have been studied in \cite{Ghosh/etc:2017}.

Let us assume that the two independent randomly censored life-time samples of size $n_1$ and $n_2$ are given
from the underlying distribution $F_{\boldsymbol{{\boldsymbol{\theta}} }_1}$ and 
$F_{\boldsymbol{{\boldsymbol{\theta}} }_2}$, respectively.
Given a general function  ${\boldsymbol{m}}({\boldsymbol{\theta}}_1, {\boldsymbol{\theta}}_2)$ 
from $\mathbb{R}^p\times\mathbb{R}^p$ to $\mathbb{R}^r$, suppose our interest is in testing 
for the general class of parametric hypotheses given by 
\begin{equation}
H_0 : {\boldsymbol{m}}({\boldsymbol{\theta}}_1, {\boldsymbol{\theta}}_2)=\boldsymbol{0}_r ~~~ \mbox{against} ~~~~ 
H_1 : {\boldsymbol{m}}({\boldsymbol{\theta}}_1, {\boldsymbol{\theta}}_2) \neq \boldsymbol{0}_r.
\label{EQ:7two_sample_Gen}
\end{equation}
The most common special case is to test for the homogeneity of the two life-time distributions 
given by ${\boldsymbol{m}}({\boldsymbol{\theta}}_1, {\boldsymbol{\theta}}_2)= ({\boldsymbol{\theta}}_1- {\boldsymbol{\theta}}_2)$
or the partial homogeneity based on a subset of (multidimensional) parameter.
As before, we need to assume that ${\boldsymbol{M}}_i({\boldsymbol{\theta}}_1, {\boldsymbol{\theta}}_2) = \frac{\partial}{\partial{\boldsymbol{\theta}}_i}{\boldsymbol{m}}({\boldsymbol{\theta}}_1, {\boldsymbol{\theta}}_2)^T$
exists, has rank $r$ and is continuous with respect to its arguments for each $i=1, 2$.

Now, let $^{(1)}\widehat{{\boldsymbol{\theta}}}_{n_1}$  and $^{(2)}\widehat{{\boldsymbol{\theta}}}_{n_2}$ denote 
the M-estimators of $\boldsymbol{\theta}_1$ and $\boldsymbol{\theta}_2$, respectively, based on each samples separately 
corresponding to a given $\boldsymbol{\psi}$-function. Also, let
$^{(i)}\widehat{\boldsymbol{\boldsymbol\Sigma}}_n(\boldsymbol\psi; ^{(i)}\widehat{\boldsymbol{{\boldsymbol\theta}}}_{n_i})$ 
be a consistent variance estimator for  $^{(i)}\widehat{\boldsymbol{{\boldsymbol\theta}}}_{n_i}$, for $i=1,2$,  from Section \ref{SEC:Var_Est}.
Then, we define the Wald-type test statistic for testing (\ref{EQ:7two_sample_Gen}) as 
\begin{equation}
\label{EQ:7_2SDT_gen}
W_{n_1, n_2}^{(2)}
= \frac{n_1n_2}{n_1+n_2} ~ {\boldsymbol{m}}\left( ^{(1)}\widehat{{\boldsymbol{\theta}}}_{n_1} , ^{(2)}\widehat{{\boldsymbol{\theta}}}_{n_2}\right)^T
\widetilde{{\boldsymbol{\Sigma}}}_{n_1, n_2}( ^{(1)}\widehat{{\boldsymbol{\theta}}}_{n_1}, ^{(2)}\widehat{{\boldsymbol{\theta}}}_{n_2})^{-1}
{\boldsymbol{m}}\left( ^{(1)}\widehat{{\boldsymbol{\theta}}}_{n_1} , ^{(2)}\widehat{{\boldsymbol{\theta}}}_{n_2}\right),
\end{equation}
where 
$\widetilde{{\boldsymbol{\Sigma}}}_{n_1, n_2}({\boldsymbol{\theta}}_1, {\boldsymbol{\theta}}_2) 
=\sum\limits_{i=1,2}\frac{n_1+n_2-n_i}{n_1+n_2} {\boldsymbol{M}}_i({\boldsymbol{\theta}}_1,{\boldsymbol{\theta}}_2)^T
 {^{(i)}}\widehat{\boldsymbol{\boldsymbol\Sigma}}_n(\boldsymbol\psi; ^{(i)}\widehat{\boldsymbol{{\boldsymbol\theta}}}_{n_i})
{\boldsymbol{M}}_i({\boldsymbol{\theta}}_1,{\boldsymbol{\theta}}_2).
$ 
We additionally need to assume that $\frac{n_i}{n_1+n_2} \rightarrow \omega_i\in(0,1)$, for both $i=1,2$, as $n_1, n_2 \rightarrow\infty$;
clearly $\omega_1 + \omega_2 =1$.
Then, it is straightforward to show from the asymptotic distribution of the M-estimators that, 
under the null hypothesis in (\ref{EQ:7two_sample_Gen}), $W_{n_1, n_2}^{(2)}$ asymptotically follows a $\chi_r^2$ distribution
as $n_1, n_2 \rightarrow\infty$.
Thus, the level-$\alpha_0$ critical region based on $W_{n_1, n_2}^{(2)}$ for testing (\ref{EQ:7two_sample_Gen}) is given by 
$\{W_{n_1, n_2}^{(2)} > \chi^2_{r,\alpha_0}\}.$

As in the case of the one-sample problem, we can also develop an approximation to the asymptotic power for 
the test based on $W_{n_1, n_2}^{(2)}$ as well, which is given by 
\begin{eqnarray}
\widetilde{\pi}_{n_1, n_2}({\boldsymbol{\theta}}_1, {\boldsymbol{\theta}}_2) = 
P\left(W_{n_1, n_2}^{(2)} > \chi_{r,\alpha}^2\right) 
\approx 1 - \Phi \left( \frac{\sqrt{\frac{n+m}{nm}}}  {2\sqrt{\widetilde{l^*}({\boldsymbol{\theta}}_1, {\boldsymbol{\theta}}_2)}} 
\left[ \chi_{r,\alpha}^2-  \frac{nm}{n+m}\widetilde{l^*}({\boldsymbol{\theta}}_1,{\boldsymbol{\theta}}_2) \right] \right), \nonumber
\end{eqnarray}
whenever ${\boldsymbol{\psi}}({\boldsymbol{\theta}}_1, {\boldsymbol{\theta}}_2)\neq \boldsymbol{0}_r$, where 
$\widetilde{l^*}({\boldsymbol{\theta}}_1,{\boldsymbol{\theta}}_2) 
= {\boldsymbol{m}}({\boldsymbol{\theta}}_1, {\boldsymbol{\theta}}_2)^T
\widetilde{{\boldsymbol{\Sigma}}}_{n_1, n_2}({\boldsymbol{\theta}}_1,{\boldsymbol{\theta}}_2)^{-1}
{\boldsymbol{m}}({\boldsymbol{\theta}}_1, {\boldsymbol{\theta}}_2).$ 
Note that, from this approximation, we also get 
$\widetilde{\pi}_{n_1, n_2}({\boldsymbol{\theta}}_1, {\boldsymbol{\theta}}_2)\rightarrow 1$ as $n_1,n_2 \rightarrow \infty$
proving the consistency of the proposed two-sample tests as well against any fixed alternative.

However, the computation of the asymptotic contiguous power is a bit tricky here,
since we can consider different possible contiguous alternatives for the two-sample case;
see the discussion in \cite{Ghosh/etc:2017}. Here, let us consider a class of general contiguous alternatives 
defined as $H_{1,n_1,n_2}: {\boldsymbol{\theta}}_i={\boldsymbol{\theta}}_{i,n} = {\boldsymbol{\theta}}_{i0} + n_i^{-\frac{1}{2}}{\boldsymbol{\Delta}}_i,
i=1,2,$
for some fixed $\left({\boldsymbol{\theta}}_{10}, {\boldsymbol{\theta}}_{20}\right)$ satisfying 
$\boldsymbol{m}({\boldsymbol{\theta}}_{1}, {\boldsymbol{\theta}}_{2})=0$
and $({\boldsymbol{\Delta}}_1, {\boldsymbol{\Delta}}_2)\in \mathbb{R}^p\times\mathbb{R}^p-\{(\boldsymbol{0}_p,\boldsymbol{0}_p)\}$. 
Then, one can show that the asymptotic distribution of $W_{n_1, n_2}^{(2)}$ under these alternatives $H_{1,m,n}$ is $\chi^2_r(\widetilde{\delta})$,
where $	\widetilde{\delta} = \boldsymbol{W}({\boldsymbol{\Delta}}_1,{\boldsymbol{\Delta}}_2)^T
	\left[\lim\limits_{n_1, n_2\rightarrow \infty}\widetilde{{\boldsymbol{\Sigma}}}_{n_1,n_2}({\boldsymbol{\theta}}_1,{\boldsymbol{\theta}}_2)\right]^{-1} 
	\boldsymbol{W}({\boldsymbol{\Delta}}_1,{\boldsymbol{\Delta}}_2)$ 
with  
$\boldsymbol{W}({\boldsymbol{\Delta}}_1,{\boldsymbol{\Delta}}_2)
=\left[\sqrt{\omega}{\boldsymbol{M}}_1^T({\boldsymbol{\theta}}_1,{\boldsymbol{\theta}}_2){\boldsymbol{\Delta}}_1 
+ \sqrt{1-\omega}{\boldsymbol{M}}_2^T({\boldsymbol{\theta}}_1,{\boldsymbol{\theta}}_2){\boldsymbol{\Delta}}_2\right].$
These asymptotic results can then be used directly to compute (or estimate) the asymptotic contiguous power of the proposed test.


Finally, in order to study the robustness of our two-sample Wald-type test statistics $W_{n_1, n_2}^{(2)}$, 
we defined the corresponding statistical functional as (ignoring the multiplier $\frac{n_1n_2}{n_1+n_2}$) 
$$
\widetilde{W}_{\boldsymbol{\psi}}^{(2)}(G_1, G_2) = {\boldsymbol{m}}^T\left(\boldsymbol{T}_{\boldsymbol{\psi}}(G_1),\boldsymbol{T}_{\boldsymbol{\psi}}(G_2)\right)
\widetilde{{\boldsymbol{\Sigma}}_{\boldsymbol{\psi}}}(\boldsymbol{T}_{\boldsymbol{\psi}}(G_1),\boldsymbol{T}_{\boldsymbol{\psi}}(G_2))^{-1}
{\boldsymbol{m}}\left(\boldsymbol{T}_{\boldsymbol{\psi}}(G_1),\boldsymbol{T}_{\boldsymbol{\psi}}(G_2)\right),
$$ 
where $G_i$ denote the true underlying distribution of the $i$-th sample for each $i=1, 2$, and 
$\widetilde{{\boldsymbol{\Sigma}}_{\boldsymbol{\psi}}}({\boldsymbol{\theta}}_1, {\boldsymbol{\theta}}_2) 
=\sum\limits_{i=1,2}\omega_i {\boldsymbol{M}}_i^T({\boldsymbol{\theta}}_1,{\boldsymbol{\theta}}_2)
{\boldsymbol{\boldsymbol\Sigma}}(\boldsymbol\psi; {\boldsymbol\theta}_i){\boldsymbol{M}}_i({\boldsymbol{\theta}}_1,{\boldsymbol{\theta}}_2).
$ 
Let us denote, under the null hypothesis in (\ref{EQ:7two_sample_Gen}), 
$G_i=F_{{\boldsymbol{\theta}}_{i0}}$ for $i=1, 2$ 
and define there contaminated versions $G_{i, \epsilon} = (1-\epsilon)G_i + \epsilon\wedge_{t_i}$, respectively, for $i=1,2$.
Note that, in two sample case, the contamination can be in either of the samples or in both of them. 
Following the similar calculations as in Section \ref{SEC:IF_test}, one can show that the first order IF 
of the two-sample test functional $\widetilde{W}_{\boldsymbol{\psi}}^{(2)}$ is again zero at the null distribution 
for either types of contaminations. The corresponding second order IFs are given by 
\begin{eqnarray}
&& IF^{(i)}_2(t_i; \widetilde{W}_{\boldsymbol{\psi}}^{(2)},  F_{{\boldsymbol{\theta}}_{10}},F_{{\boldsymbol{\theta}}_{20}}) \nonumber\\
&=& 2\mathcal{IF}(t_i; \boldsymbol{T}_{\boldsymbol{\psi}}, F_{{\boldsymbol{\theta}}_{10}})^T 
{\boldsymbol{M}}_i({\boldsymbol{\theta}}_{10},{\boldsymbol{\theta}}_{20})
\widetilde{{\boldsymbol{\Sigma}}_{\boldsymbol{\psi}}}({\boldsymbol{\theta}}_{10},{\boldsymbol{\theta}}_{20})^{-1} 
{\boldsymbol{M}}_i({\boldsymbol{\theta}}_{10},{\boldsymbol{\theta}}_{20})^T
\mathcal{IF}(t_i; \boldsymbol{T}_{\boldsymbol{\psi}}, {\boldsymbol{\theta}}_{i0}),\nonumber
\end{eqnarray}
for contamination only in the $i$-th sample ($i=1,2$) at the point $t_i$, and		
\begin{eqnarray}
IF_2((t_1, t_2); \widetilde{W}_{\boldsymbol{\psi}}^{(2)},  F_{{\boldsymbol{\theta}}_{10}},F_{{\boldsymbol{\theta}}_{20}}) 
&=& 2 \boldsymbol{Q}_{\boldsymbol{\psi}}(t_1,t_2)^T
\widetilde{{\boldsymbol{\Sigma}}_{\boldsymbol{\psi}}}({\boldsymbol{\theta}}_{10},{\boldsymbol{\theta}}_{20})^{-1}
\boldsymbol{Q}_{\boldsymbol{\psi}}(t_1,t_2),
\nonumber
\end{eqnarray}
for contamination in both the samples, where
$$\boldsymbol{Q}_{\boldsymbol{\psi}}(t_1, t_2) 
= {\boldsymbol{M}}_1^T({\boldsymbol{\theta}}_{10},{\boldsymbol{\theta}}_{20})
\mathcal{IF}(t_1; \boldsymbol{T}_{\boldsymbol{\psi}}, F_{{\boldsymbol{\theta}}_{10}})
+ {\boldsymbol{M}}_2^T({\boldsymbol{\theta}}_{10},{\boldsymbol{\theta}}_{20}) 
\mathcal{IF}(t_2; \boldsymbol{T}_{\boldsymbol{\psi}}, F_{{\boldsymbol{\theta}}_{10}}).
$$
Note that, these second order IFs are again  bounded  if the IF of the M-estimator being used is bounded 
and the later holds for all bounded $\boldsymbol{\psi}$-functions. 
Thus, the MDPDE based two sample Wald-type tests with any $\alpha>0$ yield robust solution under any types of infinitesimal contamination. 
Interestingly, for contamination in both the samples, even if the IFs of individual estimators are  not bounded, 
the proposed test can still be robust having bounded second order IF provided the term $\boldsymbol{Q}_{\boldsymbol{\psi}}(t_1, t_2)$ is bounded. 
An example of such case arises while testing for homogeneity of two samples ($H_0: \boldsymbol{\theta}_1 = \boldsymbol{\theta}_2$) 
with $t_1 = t_2$  (i.e., same contamination in both samples), 
where  $\boldsymbol{Q}_{\boldsymbol{\psi}}(t_1, t_2)$ becomes identically zero under the null.

As in the one sample case, we can also derive the level and power influence function of the proposed two-sample tests
by extending the arguments from \cite{Ghosh/etc:2017} and we omit the details for brevity. 
It can be shown that, whenever the IF of the underlying estimators are bounded,  
the LIF is again zero and PIF is a linear function (or combination) of the IFs of these estimators used.
In all case, the robustness implication turns out to be  the same as above; 
our test will be robust in both level and power when using the bounded $\boldsymbol{\psi}$-function or in particular the MDPDEs with $\alpha>0$.

\subsection{Robust Tests against One-Sided Alternatives}
\label{SEC:One-sided_problem}

Although we have developed the theory of the proposed Wald-type tests for both sided alternatives,
it can also be applied to one sided alternatives with suitable modifications when the null dimension is one. 
Such one sided alternatives commonly arise, for example, in studying the efficacy of a treatment against the control group, 
i.e., testing if the average lifetime of treatment group is greater than that of the control group. 
The main idea for testing such one sided alternative would be to consider the (suitably) signed  square-root of the 
proposed both-sided test statistics having $\chi_1^2$ asymptotic null distribution (when the dimension of the null parameter space is one);
the resulting square-root statistic then asymptotically has the standard normal null distribution
and yields robust inference against the one-sided alternatives. 

For illustration, we consider the null hypothesis in (\ref{EQ:7two_sample_Gen}) with $r=1$, 
but against the  one-sided alternative as given by 
\begin{equation}
H_0 : {{m}}({\boldsymbol{\theta}}_1, {\boldsymbol{\theta}}_2)=0 ~~~ \mbox{against} ~~~~ 
H_1 : {{m}}({\boldsymbol{\theta}}_1, {\boldsymbol{\theta}}_2) > 0.
\label{EQ:7two_sample_GenO}
\end{equation}
The modified version of the proposed test statistic for testing (\ref{EQ:7two_sample_GenO})
is then obtained from the two-sided Wald-type test statistics $W_{n_1, n_2}^{(2)}$ as
\begin{equation}
\label{EQ:7_2SDT_genO}
\widetilde{W}_{n_1, n_2}^{(2)}
= sign\left({m}\left( ^{(1)}\widehat{{\boldsymbol{\theta}}}_{\beta} , ^{(2)}\widehat{{\boldsymbol{\theta}}}_{\beta}\right)\right) \sqrt{W_{n_1, n_2}^{(2)}}
= \sqrt{\frac{n_1n_2}{n_1+n_2}} \frac{{m}\left( ^{(1)}\widehat{{\boldsymbol{\theta}}}_{n_1} , ^{(2)}\widehat{{\boldsymbol{\theta}}}_{n_2}\right)}{
	\sqrt{\widetilde{{\boldsymbol{\Sigma}}}_{n_1, n_2}( ^{(1)}\widehat{{\boldsymbol{\theta}}}_{n_1}, ^{(2)}\widehat{{\boldsymbol{\theta}}}_{n_2})}}.
\end{equation}
Note that $\widetilde{{\Sigma}_{\beta}}({\boldsymbol{\theta}}_{\beta},{\boldsymbol{\theta}}_{\beta})$ is now a scalar since $r=1$.
Clearly, under the null hypothesis in (\ref{EQ:7two_sample_GenO}), $\widetilde{W}_{n_1, n_2}^{(2)}$ asymptotically follows the standard normal distribution
and hence the level-$\alpha_0$ critical region for testing (\ref{EQ:7two_sample_GenO}) is given by 
$\left\{\widetilde{W}_{n_1, n_2}^{(2)} > z_{1-\alpha_0}\right\}$, 
where $z_{1-\alpha_0}$ denotes the $(1-\alpha)$-th quantile of the standard normal distribution.

One can further derive the detailed theory of this one-sided test statistics in a similar fashion as in the previous sections. 
In particular, we can show that this test is also consistent at any fixed alternatives and 
is robust when the $\psi$-function is chosen to be bounded. For brevity, we skip the details here
and refer the readers to \cite{Ghosh/etc:2017} where the similar theory has been developed for the MDPDE based 
one-sided tests with uncensored data; standard modifications as described previously 
will lead to the results for the present case of randomly censored data. 

\section{Real-Life Applications: MDPDE based Wald-type Tests}\label{SEC:numerical}

We now present some real-life applications of the proposed Wald-type tests based on the MDPDE
in different two-arm clinical trials including those mentioned briefly in the introduction.
Based on the nature of their cumulative hazard functions (as in Figure \ref{FIG:CumHarad}),
we model the data from each arm of a clinical trial separately as well as the combined data of both the arms 
by the Weibull distributions with parameters $(a_1, b_1)$, $(a_2, b_2)$ and $(a_0, b_0)$, respectively; 
we stick to the same notation for all trials for easier understanding.
We then apply the proposed test procedures for several hypotheses on these parameters $a_i$s and $b_i$s
under the actual data as well as after removing the outliers from the data; 
for brevity only few interesting cases are reported below.

\subsection{Veteran Lung Cancer trial}

\begin{figure}[h!t]
	\centering
	\subfloat[$H_0: b_0=1$]{
		\includegraphics[width=0.4\textwidth] {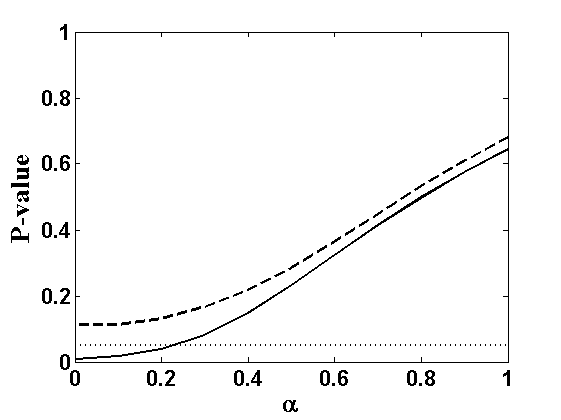}
		\label{FIG:Pvalue_Veteran_both}}
	~ 
	\subfloat[$H_0: a_2=85, b_2=1$]{
		\includegraphics[width=0.4\textwidth] {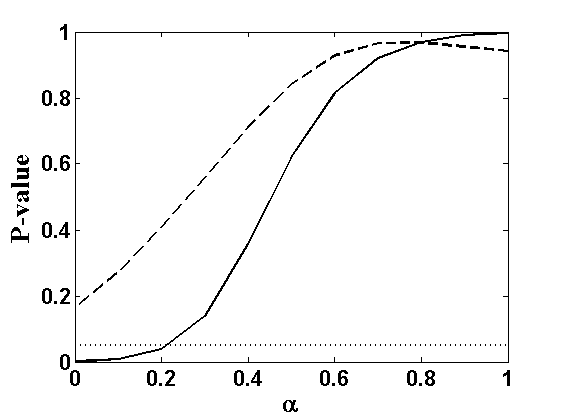}
		\label{FIG:Pvalue_Veteran_Trt2}}
	\caption{P-values for the MDPDE based Wald-type test over $\alpha$ for Veteran Lung Cancer trial 
		[Solid line: Full data; Dotted line: Outlier deleted data].}
	\label{FIG:Veteran_pvalue}
\end{figure}

We first consider the right censored data from a benchmark clinical trial study in survival analysis, 
namely the Veterans' Administration Lung Cancer study, which is a randomized trial with two treatment regimens for lung cancer. 
Arm A represents standard treatment whereas Arm B is the treatment under test; 
more details about the study can be found in \cite{Heritier/etc:2009,Kalbfleisch/Prentice:2011}.
The MDPDEs of the Weibull parameters fitted to these data are presented in Table \ref{TAB:MDPDE_4data}.
As noted earlier, the robust estimate of $b_i$ is close to one for all three cases ($i=0,1,2$),
but the MLE of $b_2$ and $b_0$ are little away from one; so we need to check if they are significantly different from one.
Note that a Weibull distribution with shape parameter $b=1$ is nothing but the simple exponential distribution
and hence a test for the hypothesis $H_0: \beta=1$ is equivalent to test for exponentiality against Weibull alternatives. 
A close investigation yields three large outlying response values (uncensored) among the 68 patients in Arm B,
which cause the  difference in the MLE.
We apply the proposed MDPDE based Wald-type tests for these data with and without these outliers;
the p-values of two particular hypotheses, namely one composite hypothesis $H_0: b_2=1$ with unknown $a_2$
and one simple hypothesis $H_0: a_2=85, b_2=1$ against their respective omnibus alternatives, 
are shown in Figure \ref{FIG:Veteran_pvalue} for different $\alpha\geq 0$.
Note that, for the classical Wald test at $\alpha=0$ the presence or absence of the outliers leads to a difference in the final conclusion, 
whereas the proposed tests with $\alpha>0.2$ give stable inference which is not affected by outliers.

Further we can apply the proposed two sample testing procedures to robustly compare the data in two arms.
For illustrations consider the problem of comparing the shape of the distributions for both arms,
which can be done by testing the composite hypothesis $H_0: b_1 = b_2$ with unknown $a_1, a_2$.
The p-value for testing this null hypothesis against the one sided alternative $H_1: b_1 > b_2$ using the classical Wald test
under the full data turns out to be 0.047 which becomes 0.23 after removing the outliers from Arm 2.
Thus, the inference at 5\% level of significance based on the classical MLE based Wald test turns around under the presence of outlying observations. 
However, the p-value resulting from the MDPDE based proposed Wald-type test for 
the above composite two-sample problems are 0.14, 0.39 and 0.56 at $\alpha=0.2, 0.5, 1$, respectively, under the full data. 
The outlier deleted data also lead to insignificant p-values in each of these cases.  
This clearly indicates the robust nature of our proposal in the two-sample problem.

\subsection{Small-Cell Lung Cancer trial}

\begin{figure}[h!t]
	\centering
	\subfloat[$H_0: a_1=770$]{
		\includegraphics[width=0.4\textwidth] {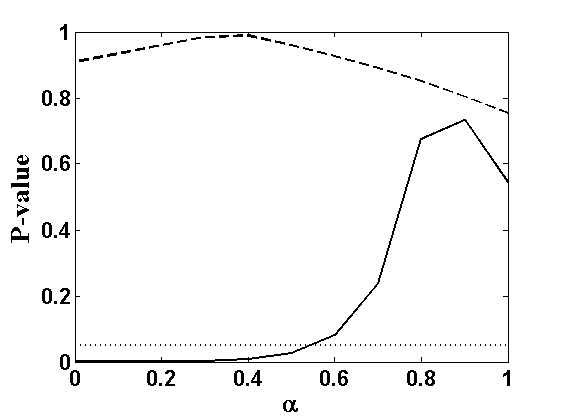}
		\label{FIG:Pvalue_SmallCell_Trt0}}
	~ 
	\subfloat[$H_0: b_2=2$]{
		\includegraphics[width=0.4\textwidth] {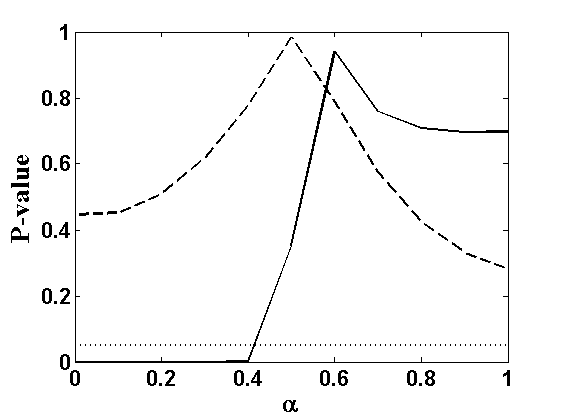}
		\label{FIG:Pvalue_SmallCell_Trt1}}
	\caption{P-values for the MDPDE based Wald-type test over $\alpha$ for Small-Cell Lung Cancer  trial 
		[Solid line: Full data; Dotted line: Outlier deleted data].}
	\label{FIG:SmallCell_pvalue_ArmA}
\end{figure}

We now consider randomly right-censored survival time data from another two arm clinical trial
on the Small-cell lung cancer with two different treatments \citep{Ying/etc:1995}.
The data on both the arms contain few outlying observations which make the MLE different from 
the robust MDPDE of the Weibull parameters as presented in Table \ref{TAB:MDPDE_4data}.
We again apply the proposed Wald-type tests based on these MDPDEs under the full data 
as well as the outlier deleted data. The resulting p-values are shown in Figure \ref{FIG:SmallCell_pvalue_ArmA}
for the composite hypotheses $H_0: a_1=770$ with unknown $b_1$ and $H_0: b_2=2$ with unknown $a_2$, against their respective omnibus alternatives.
Once again, it is pretty clear that the proposed tests produce more stable inference than the classical Wald test
but we need slightly larger $\alpha$ ($\geq 0.5$ or thereabouts) due to the presence of heavy outliers.

\begin{figure}[h!t]
	\centering
	\subfloat[$H_0: a_0=2920, b_0=1.3$]{
		\includegraphics[width=0.4\textwidth] {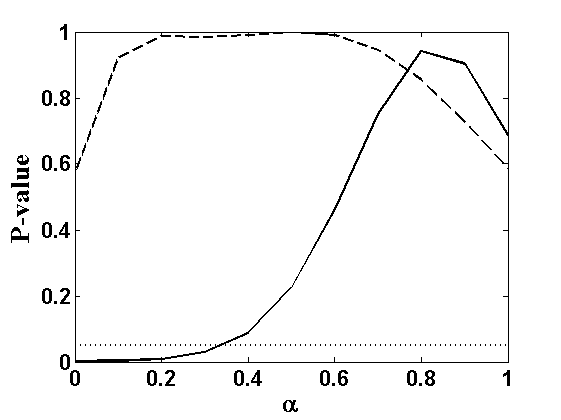}
		\label{FIG:Pvalue_BRCAOV_both}}
	~ 
	\subfloat[$H_0: b_2=1.45$]{
		\includegraphics[width=0.4\textwidth] {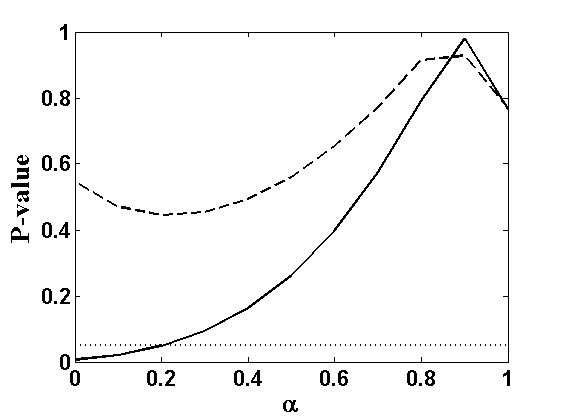}
		\label{FIG:Pvalue_BRCAOV_Trt2}}
	\caption{P-values for the MDPDE based Wald-type test over $\alpha$ for Breast \& Ovarian Cancer	trial 
		[Solid line: Full data; Dotted line: Outlier deleted data].}
	\label{FIG:BRCAOV_pvalue_ArmA}
\end{figure}

\subsection{Breast and Ovarian Cancer trial}

Our third example is from a Breast and Ovarian cancer (BOC) trial where the two arms contain the randomly censored survival times 
of the breast cancer patients  and the Ovarian cancer patients, respectively. 
The data are available in the recent R-package \textit{survminer} and the parameter estimates
under the fitted Weibull model are presented in Table \ref{TAB:MDPDE_4data}.
Again some large outlying observations affect the MLE significantly. Motivated by their robust estimates,
we perform the proposed MDPDE based test for the simple hypothesis $H_0: a_0=2920, b_0=1.3$
and the composite hypothesis $H_0: b_2=1.45$ with unknown $a_2$, against their respective omnibus alternatives
and plot the resulting p-values in Figure \ref{FIG:BRCAOV_pvalue_ArmA} for both the full data 
and outlier deleted data. The proposed tests with slightly larger $\alpha>0$ are again seen 
to produce robust inference in contrast to the non-robust classical Wald test (at $\alpha=0$).

\begin{figure}[h!t]
	\centering
	\subfloat[$H_1: b_1\neq 1$]{
		\includegraphics[width=0.4\textwidth] {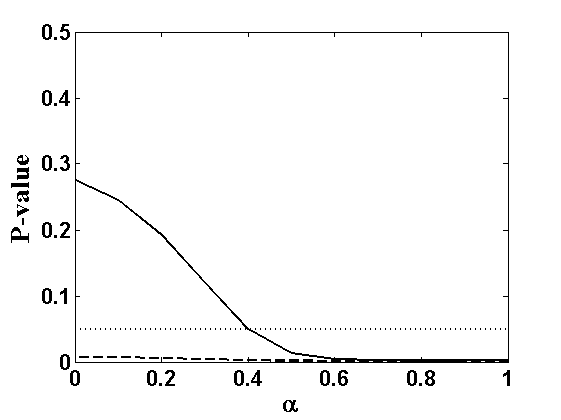}
		\label{FIG:Pvalue_Gastric_2tail}}
	~ 
	\subfloat[$H_1: b_1< 1$]{
		\includegraphics[width=0.4\textwidth] {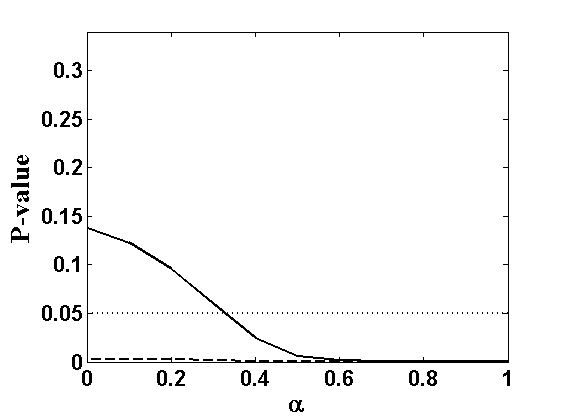}
		\label{FIG:Pvalue_Gastric_1tail}}
	\caption{P-values for the MDPDE based Wald-type test over $\alpha$ for testing exponentiality of Arm A (i.e., $H_0: b_1=1$) 
		of Gastric Carcinoma trial against two alternatives
		[Solid line: Full data; Dotted line: Without two early failures].}
	\label{FIG:Gastric_pvalue_ArmA}
\end{figure}

\subsection{Gastric Carcinoma trial}

Our next example is based on another two arm clinical trial data on 90 locally unresectable gastric cancer patients
randomized into two equal arms \cite{Klein/Moeschberger:2003}.
Patients in arms A and B are, respectively, treated with only chemotherapy and a combined treatment of chemotherapy and radiotherapy.
The data were initially produced by the Gastrointestinal Tumor Study Group  \cite{Gastric:1982} and later analyzed by many scientists 
\cite[see, e.g.,][]{Stablein/Koutrouvelis:1985}.  Interestingly, this study is different from the previous examples
in that the data in both the arms contain some early failures, a different type of outliers,  
which also affects the MLE significantly and forces it to be different from the robust parameter estimates
(Table \ref{TAB:MDPDE_4data}). Suppose we want to test for the exponentiality of the data in Arm A, i.e., 
to test for the composite null hypothesis $H_0: b_1 =1$ assuming $a_1$ to be unknown.
We apply the proposed MDPDE based tests against the both sided alternative $H_1: b_1\neq 1$ and also against the one sided alternative
$H_1: b_1 <1$; the resulting p-values are reported in Figure \ref{FIG:Gastric_pvalue_ArmA} 
based on the full data as well as after removing two early failures from Arm A. 
Note that all the estimates of $b_1$ in Table \ref{TAB:MDPDE_4data} are larger than one, 
but the MLE based classical Wald test fails to reject the null hypothesis in favor of either alternative in presence of the early failures.
The proposed MDPDE based tests, on the other hand, produce robust inference at $\alpha \geq 0.4$ 
even under the presence of such early failure observations.

\subsection{Head and Neck Cancer trial}

We now present another analysis of a clinical trial data for head and neck cancer treatment \cite{Efron:1988},
which has earlier been used to illustrate the advantages of MDPDE over MLE in \cite{Basu/etc:2006}.
The two arms of the dataset represent the radiation therapy alone (arm A) and a  combination of radiation therapy and chemotherapy (arm B), respectively.
Here, we restrict our attention to arm A, which contains seven (7) large outliers among a total of $51$.
Further, nine (9) patients in this arm were lost to follow-up, producing a high censoring rate of about 20\%.
As demonstrated in \cite{Basu/etc:2006}, the robust MDPDE of the Weibull parameter $(a_1, b_1)$ is approximately $(250, 1.47)$ at $\alpha=1$, 
which fit the majority of the data quite well, compared to the non-robust AMLEs $(418, 0.98)$; 
see Table \ref{TAB:MDPDE_2data} for MDPDEs at other $\alpha$.
Further, after deleting the 7 outliers from the data, the corresponding AMLE becomes approximately $(239, 1.46)$ 
which is quite close to the robust  MDPDEs derived from the full data including outliers.

\begin{figure}[h!t]
	\centering
	\subfloat[$H_0: a_1=250, b_1=1.4$]{
		\includegraphics[width=0.4\textwidth] {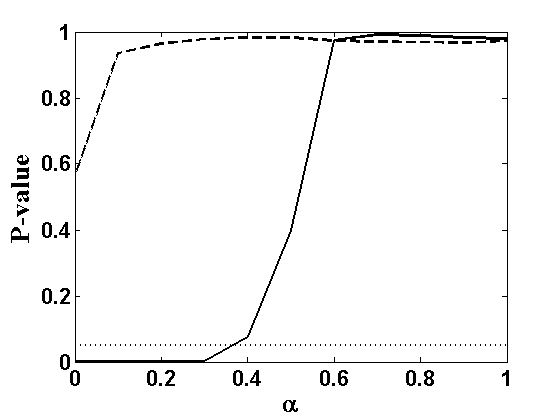}
		\label{FIG:Efron_pvalue_1}}
	~ 
	\subfloat[$H_0: a_1 = 418$]{
		\includegraphics[width=0.4\textwidth] {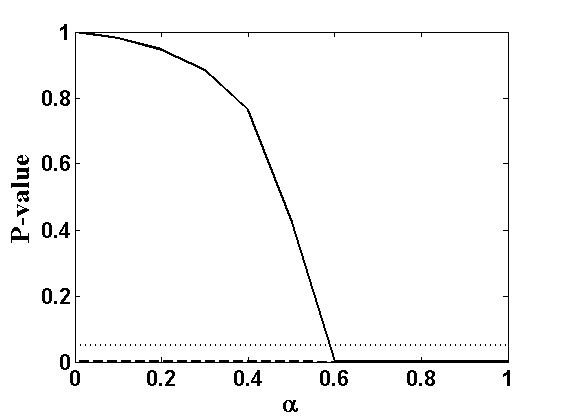}
		\label{FIG:Efron_pvalue_2}}
	\caption{P-values for the MDPDE based Wald-type test over $\alpha$ for Head and Neck Cancer trial
		[Solid line: Full data; Dotted line: Outlier deleted data].}
	\label{FIG:Efron_pvalue_ArmA}
\end{figure}

Motivated by the above analyses, we present the p-values obtained by the proposed MDPDE based tests
of two particular hypotheses in Figure \ref{FIG:Efron_pvalue_ArmA}; 
one simple hypothesis $H_0: a_1=250, b_1=1.4$, and one composite hypothesis $H_0: a_1 = 418$ with unknown $b_1$,
against their respective omnibus alternatives. 
Note that, due to the presence of outliers, the AMLE based classical Wald-test soundly produces incorrect inferences, 
as compared to the results obtained after removing the outliers. 
But, the proposed Wald-type tests can successfully ignore the effect of outliers 
and produce stable results for all $\alpha\geq 0.4$.

\begin{table}[h]
\centering
\caption{MDPDEs of ($a, b$) in the Weibull model fitted to two datasets}
\begin{tabular}{l| cccccc} \hline
Parameter		&  		\multicolumn{6}{c}{$\alpha$}\\
& 0 (AMLE) &	0.1	&	0.3 & 0.5 & 0.7 & 1	\\\hline
\multicolumn{7}{l}{\underline{Head and Neck Cancer data}}\\
$a$ & 417.98	& 412.42	&	386.62	&	321.25	&	254.92	&	249.52\\
$b$ & 0.98	& 0.99	&	1.03	&	1.16	&	1.42	& 1.47\\\hline
\multicolumn{7}{l}{\underline{Myeloma data}}\\
$a$ & 32.94	& 33.06	& 33.19	&	32.88	&	32.10	&		30.46\\
$b$ & 1.06	& 1.05	&	1.03	&	1.02	&	1.03	&	1.06\\

\hline
\end{tabular}
\label{TAB:MDPDE_2data}
\end{table}

\subsection{Myeloma Data}

\begin{figure}[h!t]
	\centering
	\subfloat[$H_0: a=30, b=1$]{
		\includegraphics[width=0.4\textwidth] {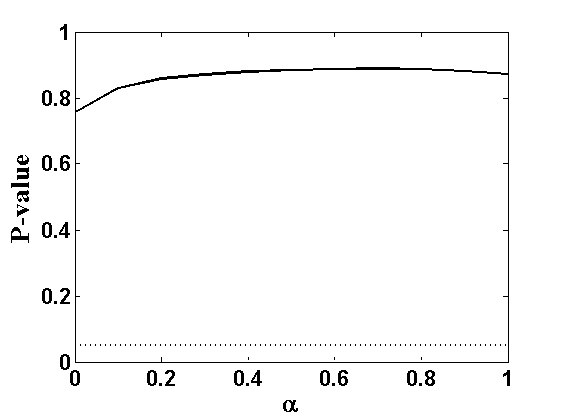}
		\label{FIG:Efron_pvalue_ArmA_H0}}
	~ 
	\subfloat[$H_0: b=1$]{
		\includegraphics[width=0.4\textwidth] {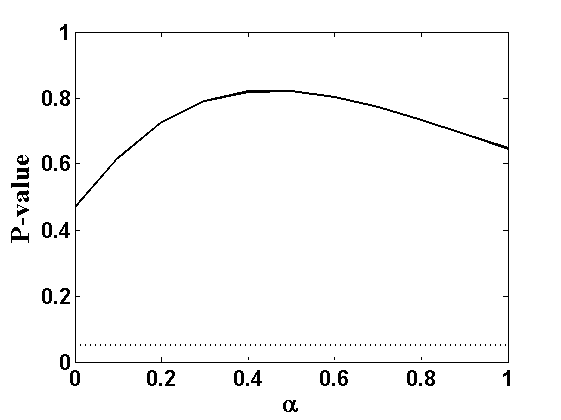}
		\label{FIG:Efron_pvalue_ArmA_H0p}}
	\caption{P-values for the MDPDE based Wald-type test over $\alpha$ for Myeloma data 
		[Solid line: Full data; Dotted line: Outlier deleted data].}
	\label{FIG:Myeloma_pvalue_ArmA}
\end{figure}

We finally present a different medical data example containing no outliers to illustrate the equally effective performance of our proposed tests.
The data consist of the lifetimes of 65 multiple myeloma patients from their diagnosis, of which seventeen (17) were still alive 
at the end of the study and hence censored \citep{Heritier/etc:2009,Krall/ect:1975}.
We fit the Weibull$(a, b)$ model to these data; the AMLE and MDPDEs are presented in Table \ref{TAB:MDPDE_2data}.
Note that, all estimates are pretty close to each other due to the absence of outliers.  
Based on these estimates, in Figure \ref{FIG:Myeloma_pvalue_ArmA}, we have plotted the p-values obtained by the proposed MDPDE based
Wald-type tests for the hypotheses $H_0: a=30, b=1$ and $H_0: b=1$ with unknown $a$, against their respective omnibus alternatives.
Clearly the p-values are also very similar to the classical test for any $\alpha \geq 0$.
This is highly important as we do not often know anything about the presence or the extent of outliers in our real-life data. 
This example illustrates that the proposed robust test can be quite similar to the classical Wald test in the absence of outliers 
with all the methods giving similar inference.

\section{Concluding Remarks}\label{SEC:conclusion}

In this paper, we have considered the fully parametric robust inference for survival data with random censoring.
We have proposed robust Wald-type tests for both simple and composite null hypotheses 
using the M-estimators including the MDPDEs; for this purpose, 
a consistent estimator for their asymptotic variance is developed. 
We have also derived the asymptotic and robustness theory of the proposed Wald-type tests based on general M-estimators.
The robustness of the proposed Wald-type tests are also studied theoretically 
through IF of the test statistics, LIF and PIF. 
Extensions for general two sample problems and one-sided alternatives are also discussed along with several real-life applications.
A natural follow up would include robust Wald-type tests for the regression set-up with randomly censored observations,
which we hope to take up in the future.

\appendix
\section{Appendix: Assumptions Required for Asymptotic Results}\label{APP:Assumptions}
\noindent
\textbf{Assumptions related to General M-estimators:}
For any distribution function $F$, let us define $\Delta F(x) = F(x) - F(x-)$
and denote the upper bound of the support  $F$ by $\tau_F = \sup\{x: F(x)<1\}$.
\begin{itemize}
	\item[(A1)] Either (i)  for each $j$, there exists some $b<\tau_{G_Z}$ 
	such that $\psi_j(x;{\boldsymbol\theta}_0)=0$ for $b< x \leq \tau_{G_Z}$; \\
	or, (ii) $\tau_{G_X} \leq \tau_{G_C}$, with strict inequality when $G_C$ is continuous at $\tau_{G_X}$
	and  $\Delta G_X(\tau_{G_X}) >0$.

	\item[(A2)] For each $j=1,\ldots,p$, $\psi_j(x;{\boldsymbol\theta}_0)$ satisfies 
	$E\left[\psi_j(Z;{\boldsymbol\theta}_0)\gamma_0(Z)\delta\right]^2 
	= \int \psi_j^2(z;{\boldsymbol\theta}_0)\gamma_0^2(z)dG_{Z,1}(z) < \infty$
	and $\int \left|\psi_j(x;{\boldsymbol\theta}_0)\right|\gamma^{1/2}(x)dG_X(x) < \infty$.
	
	\item[(A3)] ${\boldsymbol\theta}_0$ is the unique root of (\ref{EQ:est-eqn-psiX}).

	\item[(A4)] The $p\times p$ matrix $\boldsymbol{C}(\boldsymbol\psi;{\boldsymbol\theta}_0)$, 
	defined in Section \ref{SEC:Mest}, is finite element-wise.

	\item[(A5)] The $p\times p$ matrix $\boldsymbol\Lambda(\boldsymbol\psi;{\boldsymbol\theta}_0)$,
	defined in Section \ref{SEC:Mest}, is finite and non-singular. 
	
	\item[(A6)] For each $i,j=1,\ldots,p$, $g(x,{\boldsymbol\theta})= \frac{\partial}{\partial{\boldsymbol\theta}_i}\psi_j(x;{\boldsymbol\theta}_0)$ 
	is absolutely integrable with respect to $G_X$ and satisfies any one of the following conditions:
	\begin{itemize}
		\item[(i)] $g(x,{\boldsymbol\theta})$ is continuous at ${\boldsymbol\theta}_0$ uniformly in $x$,
		\item[(ii)] $\int \sup_{{\boldsymbol\theta}: \left|{\boldsymbol\theta}-{\boldsymbol\theta}_0\right|\leq \delta}
		\left|g(x,{\boldsymbol\theta}) - g(x;{\boldsymbol\theta}_0)\right|dG_X(x) =h_\delta \rightarrow 0$ as $\delta \rightarrow 0$,
		\item[(iii)] $g(x,{\boldsymbol\theta})$ is continuous in $x$ for ${\boldsymbol\theta}$ in a neighborhood of ${\boldsymbol\theta}_0$, and 
		$\lim\limits_{{\boldsymbol\theta}\rightarrow{\boldsymbol\theta}_0}
		\left|\left|g(\cdot,{\boldsymbol\theta}) - g(\cdot,{\boldsymbol\theta}_0)\right|\right|_v = 0$,
		\item[(iv)] $\int g(x,{\boldsymbol\theta})dG_X(x)$ is continuous at ${\boldsymbol\theta}= {\boldsymbol\theta}_0$, 
		and $g(x,{\boldsymbol\theta})$ is continuous in $x$ for ${\boldsymbol\theta}$ in a neighborhood of ${\boldsymbol\theta}_0$, and 
		$\lim\limits_{{\boldsymbol\theta}\rightarrow{\boldsymbol\theta}_0}\left|\left|g(\cdot,{\boldsymbol\theta}) - g(\cdot,{\boldsymbol\theta}_0)\right|\right|_v < \infty$,
		\item[(v)] $\int g(x,{\boldsymbol\theta})dG_X(x)$ is continuous at ${\boldsymbol\theta}= {\boldsymbol\theta}_0$, 
		and $\int g(x,{\boldsymbol\theta})d\widehat{G_X}(x) \mathop{\rightarrow}^\mathcal{P} \int g(x,{\boldsymbol\theta}_0)dG_X(x) < \infty$,
		uniformly for ${\boldsymbol\theta}$ in a neighborhood of ${\boldsymbol\theta}_0$.
	\end{itemize}
	
	\item[(A7)] There exists a compact set $K\subseteq\mathbb{R}^p$ such that, for each $j$,
	$
	\inf_{{\boldsymbol\theta}\notin K} \left|\int \psi_j(x;{\boldsymbol\theta})dG_X(x)\right| >0.
	$
\end{itemize}

\noindent
\textbf{Assumptions related to MDPDEs:}
\begin{itemize}
	\item[(B1)] The model distribution $F_{\boldsymbol{\theta}}$ has support independent of $\boldsymbol\theta$
	which is the same as that of $G_X$. 
	
	\item[(B2)] There is an open subset $\omega\subseteq \Theta$ containing the true parameter $\boldsymbol\theta_0$
	such that, for all $\boldsymbol\theta\in \omega$, $\int f_{\boldsymbol{\theta}}^{1+\alpha} <\infty$ 
	and $f_{\boldsymbol{\theta}}(x)$ is thrice continuously differentiable in $\boldsymbol{\theta}$ 
	for almost all $x$ in its support.
	
	\item[(B3)] $\int f_{\boldsymbol{\theta}}(x)^{1+\alpha}dx$ and
	$\int f_{\boldsymbol{\theta}}(x)^{\alpha}dG_X(x)$ are thrice differentiable and 
	the derivatives can be interchanged with the integral. 
	Also, $E_{G_X}\left[\frac{\partial V_{\boldsymbol{\theta}}(X)}{\partial\boldsymbol{\theta}}\right]<\infty$, 
	where
	$
	V_{\boldsymbol{\theta}}(x) = \int f_{\boldsymbol{\boldsymbol\theta}}^{1+\alpha}(y)dy 
	- \frac{1+\alpha}{\alpha} f_{\boldsymbol{\boldsymbol\theta}}^{\alpha}(x). 
	$
	
	\item[(B4)] The matrix $\boldsymbol\Lambda(\boldsymbol\psi_\alpha; {\boldsymbol\theta})$ 
	has all entries finite and is positive definite.
	
	\item[(B5)] For all  $\boldsymbol{\theta} \in \omega$, every third derivatives of $V_{\boldsymbol{\theta}}(x)$ 
	with respect to $\boldsymbol{\theta}$ is bounded by some function of $x$ not depending on $\boldsymbol{\theta}$
	and having finite expectation with respect to $G_X$.
\end{itemize}


\end{document}